\documentclass[aps,prd,nofootinbib,twocolumn,superscriptaddress,preprintnumbers,balancelastpage,longbibliography]{revtex4-1}

\usepackage{placeins}
\usepackage{amsmath,amssymb,mathtools,bm}
\usepackage{lipsum}
\usepackage{comment}
\usepackage{gensymb}
\usepackage{graphicx, color, hepunits}
\usepackage[dvipsnames]{xcolor}
\usepackage{float}
\usepackage{filecontents}
\usepackage{multirow}
\usepackage{slashed}
\usepackage{pifont}
\usepackage[colorlinks=true 
,urlcolor=purple 
,anchorcolor=blue
,citecolor=blue 
,filecolor=magenta  
,linkcolor=blue 
,menucolor=blue
,linktocpage=true
,pdfproducer=medialab
,pdfa=true
]{hyperref}

\usepackage[utf8]{inputenc}
\usepackage[english]{babel}
\usepackage{soul}
\usepackage{hyperref}

\newcommand{\es}[2] {\begin{equation} \label{#1} \begin{split} #2 \end{split} \end{equation}}

\makeatletter
\newcounter{savesection}
\newcounter{apdxsection}
\renewcommand\appendix{\par
  \setcounter{savesection}{\value{section}}%
  \setcounter{section}{\value{apdxsection}}%
  \setcounter{subsection}{0}%
  \gdef\thesection{\@Alph\c@section}}
\newcommand\unappendix{\par
  \setcounter{apdxsection}{\value{section}}%
  \setcounter{section}{\value{savesection}}%
  \setcounter{subsection}{0}%
  \gdef\thesection{\@arabic\c@section}}
\makeatother

\begin{document}

\title{
Time-delayed gamma-ray signatures of heavy axions from core-collapse supernovae
}

\author{Joshua N. Benabou}
\affiliation{Theoretical Physics Group, Lawrence Berkeley National Laboratory, Berkeley, CA 94720, U.S.A.}
\affiliation{Berkeley Center for Theoretical Physics, University of California, Berkeley, CA 94720, U.S.A.}

\author{Claudio Andrea Manzari}
\affiliation{Theoretical Physics Group, Lawrence Berkeley National Laboratory, Berkeley, CA 94720, U.S.A.}
\affiliation{Berkeley Center for Theoretical Physics, University of California, Berkeley, CA 94720, U.S.A.}

\author{Yujin Park}
\affiliation{Theoretical Physics Group, Lawrence Berkeley National Laboratory, Berkeley, CA 94720, U.S.A.}
\affiliation{Berkeley Center for Theoretical Physics, University of California, Berkeley, CA 94720, U.S.A.}

\author{Garima Prabhakar}
\affiliation{Berkeley Center for Theoretical Physics, University of California, Berkeley, CA 94720, U.S.A.}

\author{Benjamin R. Safdi}
\affiliation{Theoretical Physics Group, Lawrence Berkeley National Laboratory, Berkeley, CA 94720, U.S.A.}
\affiliation{Berkeley Center for Theoretical Physics, University of California, Berkeley, CA 94720, U.S.A.}

\author{Inbar Savoray}
\affiliation{Theoretical Physics Group, Lawrence Berkeley National Laboratory, Berkeley, CA 94720, U.S.A.}
\affiliation{Berkeley Center for Theoretical Physics, University of California, Berkeley, CA 94720, U.S.A.}

\date{\today}

\begin{abstract}
Heavy axions that couple to both quantum electrodynamics and quantum chromodynamics with masses on the order of MeV -- GeV and high-scale decay constants in excess of $\sim$$10^8$ GeV 
may arise generically in {\it e.g.} axiverse constructions.  In this work we provide the most sensitive search to-date for the existence of such heavy axions using Fermi-LAT data towards four recent supernovae (SN): Cas A, SN1987A, SN2023ixf, and SN2024ggi. We account for heavy axion production in the proto-neutron-star cores through nuclear and electromagnetic processes and then the subsequent decay of the axions into photons. While previous works have searched for gamma-rays from SN1987A using the Solar Maximum Mission that observed SN1987A during the SN itself, we show that using Fermi Large Area Telescope data provides an approximately five orders of magnitude improvement in flux sensitivity for axions with lifetimes larger than around 10 years. 
We find no evidence for heavy axions and exclude large regions of previously-unexplored parameter space. 
\end{abstract}
\maketitle

\begin{figure}[!htb]
    \centering
    \includegraphics[width=0.48\textwidth]{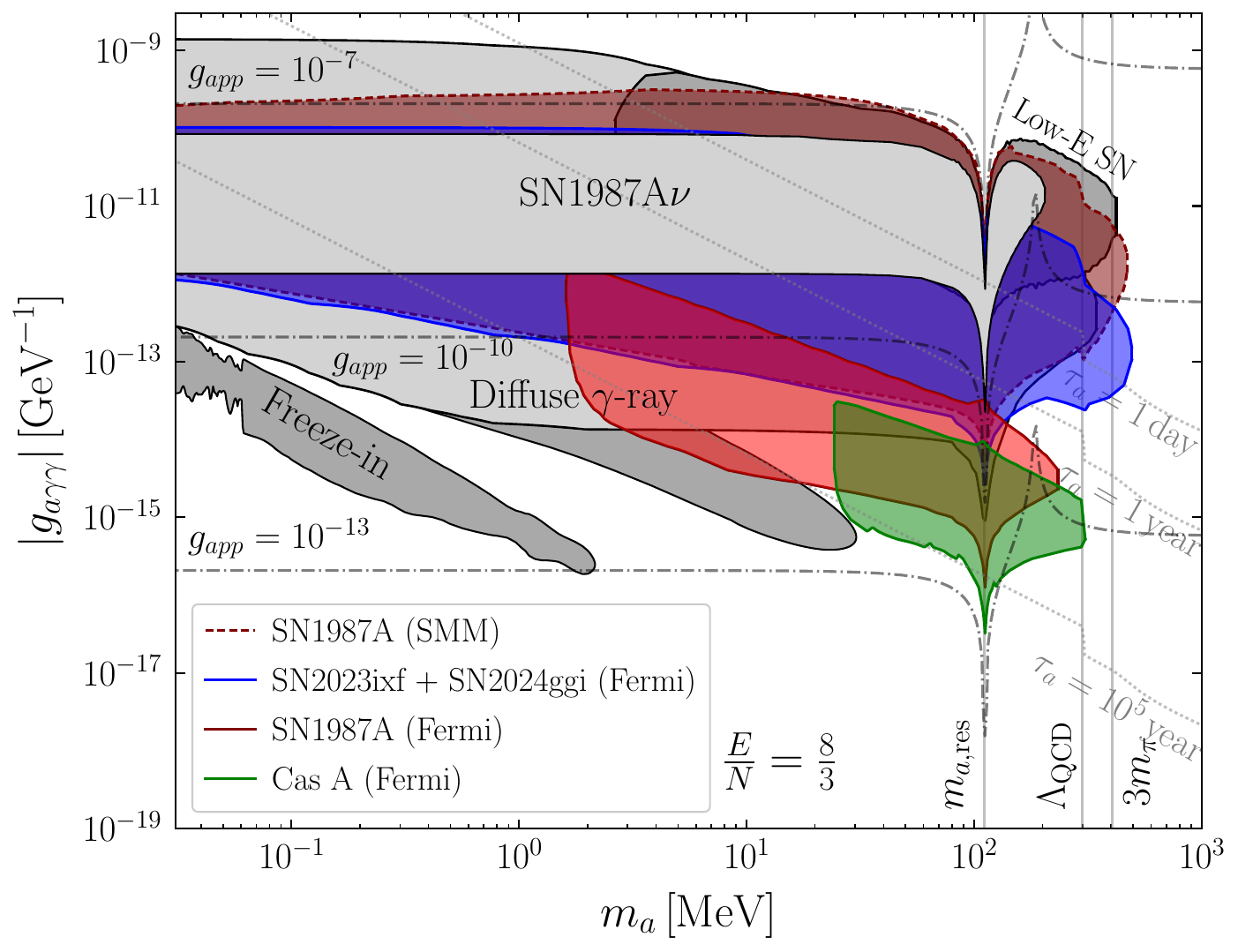}
        \caption{ 
        Excluded regions found in this work (at 95\% confidence or greater) in the $m_a$-$g_{a\gamma\gamma}$ parameter space for heavy axions that couple to QED (with anomaly coefficient $E$) and QCD (anomaly coefficient $N$) with $E/N = 8/3$, as in GUT models, from gamma-ray observations with the Fermi-LAT towards SN1987A, Cas A, SN2023ixf, and SN2024ggi, along with SMM gamma-ray observations of SN1987A.  (Note that the SN2023ixf and SN2024ggi results are joined together.) 
        We find no evidence for an axion-induced signal. 
    Additionally, we show in light gray 
    constraints we derive from comparing the diffuse SN axion-induced gamma-ray background to Fermi data and from the cooling of SN1987A (see the SM).
    In dark gray we show the irreducible axion background (`Freeze-in') constraint~\cite{Langhoff:2022bij}. 
     In dashed gray (black) we indicate contours of constant axion lifetime (axion-proton coupling); see also SM Fig.~\ref{fig:gapp_constraints} for the constraints illustrated in terms of $g_{app}$ and $f_a$.  We also show the mass $m_{a, {\rm res}}$ where $g_{a\gamma\gamma}$ vanishes, $\Lambda_{\rm QCD} \approx 300$ MeV, below which axion-pion mixing occurs, and $3 m_\pi$, above which the axions can decay to three pions. 
    }
    \label{fig:gayy_constraints}
\end{figure}

\textbf{\textit{Introduction}}
Heavy axions are motivated extensions to the Standard Model that 
are closely related to the quantum chromodynamics (QCD) axion, which may solve the strong-{\it CP} problem and explain dark matter (DM)~\cite{Peccei:1977hh,Peccei:1977ur,Weinberg:1977ma,Wilczek:1977pj,Preskill:1982cy,Abbott:1982af,Dine:1982ah}.  The QCD axion is an ultra-light pseudo-scalar whose small mass is generated by non-perturbative QCD effects.  Heavy axions would interact with the Standard Model through the same operators as the QCD axion but would acquire their dominant mass contributions by interactions with other sectors, such as instantons of non-abelian dark sectors or through string theory instantons such as Euclidean D-branes.  String axiverse constructions, for example, argue for a nearly log-flat distribution of axion masses including the $\sim$MeV -- GeV masses considered in this work~\cite{Green:1984sg,Witten:1984dg,Svrcek:2006yi,Arvanitaki:2009fg,Halverson:2019cmy,Conlon:2006tq,Acharya:2010zx,Ringwald:2012cu,Cicoli:2012sz,Demirtas:2021gsq,Mehta:2021pwf,Gendler:2023kjt}.   
Heavy axions may leave unique, observable signatures in a variety of astrophysical and cosmological probes. (See Fig.~\ref{fig:gayy_constraints}.)

In this work we probe new regions of heavy axion parameter space in the MeV to GeV mass range for axions that couple to quantum electrodynamics (QED) and QCD through operators of the form
\begin{equation}
\label{eq:axion_EFT}
\mathcal{L}\supset\frac{a}{F_a}\left[\frac{\alpha_{\mathrm{em}}}{8 \pi} E F_{ \mu \nu} \widetilde{F}^{\mu \nu}+\frac{\alpha_s}{8 \pi} N G_{\mu \nu}^a \widetilde{G}^{a, \mu \nu}\right] \,,
\end{equation}
with $E$ $(N)$ the electromagnetic (color) anomaly coefficient, $\alpha_{\rm EM}$ ($\alpha_s$) the electromagnetic (QCD) fine structure constant, $F$ ($G$) the QED (QCD) field strength, and $2 \pi F_a$ the periodicity of the axion field from hidden-sector instantons (not shown), which dominate the axion's potential.
While such heavy axions that couple to both QCD and QED do not solve the strong-{\it CP} problem, they allow for a standard, lighter QCD axion that does (see the Supplementary Materials (SM)).
We search for these axions using Fermi Large Area Telescope (LAT) gamma-ray data towards the past supernovae (SN) Cas A and SN 1987A  
along with more recent  
 SN2023ixf and SN2024ggi.    
 Our work builds upon previous searches for heavy axion production and decay from SN1987A~\cite{Jaeckel:2017tud,Hoof:2022xbe,Muller:2023vjm,Lella:2024dmx}. 
 Crucially, however, prior works only make use of data from the Solar Maximum Mission (SMM), which famously collected data in the direction of SN1987A (through its side shielding) when the SN exploded but did not observe a gamma-ray burst~\cite{Chupp:1989kx,Oberauer:1993yr}. On the other hand, as we show in this work much stronger constraints on heavy axions are obtained using Fermi-LAT data towards SN1987A from 2008 onwards. The reason is that the Fermi-LAT, which is a far superior instrument relative to the SMM, is able to achieve approximately five orders of magnitude stronger flux limits on the $\sim$100 MeV flux from SN1987A relative to the SMM and is able to probe axions with lifetimes much longer than 10 years. While SN2023ixf has been considered before for heavy axions~\cite{Muller:2023pip} in the context of Fermi-LAT data, we are able to achieve superior sensitivity to axions relative to that work both by accounting for axion-nucleon production and by using a more appropriate set of data quality cuts, which removes crucial instrumental gaps that limited the analysis in~\cite{Muller:2023pip}. We provide the first analysis of SN2024ggi using Fermi-LAT and find no significant evidence for axions for that or any of the other SN.

 Our work complements the search for QCD and ultralight axions using gamma-ray data towards SN including SN1987A~\cite{Raffelt:1996wa,Brockway:1996yr,Grifols:1996id,Payez:2014xsa,Hoof:2022xbe,Caputo:2024oqc,Manzari:2024jns}.  These searches account for the same axion production mechanisms considered here but 
 then the conversion of those axions to photons on Galactic and stellar magnetic fields.  

 \textbf{\textit{Heavy axion theory}} While most works have only allowed heavy axions to have couplings to electromagnetism~\cite{Jaeckel:2017tud,Hoof:2022xbe,Muller:2023vjm} (but see~\cite{Giannotti:2010ty,Lella:2024dmx}), here we argue that generically we expect heavy axions to couple both to QCD and to QED with roughly equal strengths (see~\eqref{eq:axion_EFT}).  Allowing for QCD couplings, as we show, qualitatively changes how one goes about searching for the axion decays because generically we are able to probe much higher decay constants in this case, which corresponds to longer axion lifetimes.

The axion-photon coupling in the infrared (IR) below the QCD confinement scale, for $m_a \ll m_\pi$ with $m_\pi$ the pion mass, is~\cite{diCortona:2015ldu}
\begin{align}
\label{eq:gagg}
g_{a \gamma \gamma}\approx\frac{\alpha_{\mathrm{em}}}{2 \pi f_a}\left(\frac{E}{N}-1.92\right) \,,
\end{align}
where \mbox{$f_a \equiv F_a / N$} is the axion decay constant; this equation receives substantial corrections for $m_a$ near and above the pion mass, as we discuss in the SM.  In addition to the axion couplings to gauge bosons in~\eqref{eq:axion_EFT}, the axion could also have UV derivative couplings to SM fermion axial currents. We do not consider such contributions here to be conservative. In the IR the axion acquires derivative couplings to nucleons,
\begin{align}
{\mathcal L} \supset  \frac{\partial_\mu a}{2f_a} \Bigl[
 C_{app} \bar p \gamma^\mu\gamma^5 p  +C_{ann} \bar n \gamma^\mu\gamma^5 n
\Bigr] \,,
\label{eq:Lagnucleons}
\end{align}
due to the axion coupling to $G \tilde G$. In particular, the dimensionless coefficients that appear above are given by $C_{app} \approx -0.50$ and $C_{ann} \approx -0.02$ in the limit $m_a \ll m_\pi$~\cite{GrillidiCortona:2015jxo}; we define dimensionless couplings such as $g_{app} = m_p C_{app} / f_a$ with $m_p$ the proton mass, see Fig.~\ref{fig:gayy_constraints}. 
(See the SM for the couplings at $m_a \sim m_\pi$.) In addition, the axion acquires axion-nucleon-pion and axion-nucleon-$\Delta$ interactions (see, {\it e.g.},~\cite{Chang:1993gm,Vonk:2021sit,Manzari:2024jns}).

If the Standard Model unifies into a grand unified theory (GUT), then the axion necessarily must have non-trivial $N$ and $E$~\cite{Agrawal:2022lsp,Agrawal:2024ejr}. More precisely, let us assume that the axion is neutral under the GUT gauge group and a ``standard'' embedding of the Standard Model in the GUT theory, {\it e.g.} for which the Standard Model arises from an $\mathrm{SU}(5)$ subgroup of the potentially larger group $G$. This includes an $\mathrm{SU}(5)$ GUT  with $\overline{\mathbf{5}}+\mathbf{1 0}$ matter content, $\mathrm{SO}(10)$ with $\mathbf{1 6}$ matter, and $E_6$ with $\mathbf{27}$ matter. In this case, $E/N=8/3$ \cite{Agrawal:2022lsp}.  We use $E/N = 8/3$ as a benchmark throughout this work.

In contrast to this work, much effort is focused at the moment on searching for ultra-light axions (with masses much less than the QCD-induced axion mass $m_{a,{\rm QCD}}$) that couple to QED ({\it i.e.}, have $E \neq 0$) -- see, {\it e.g.},~\cite{Safdi:2022xkm,OHare:2024nmr,Carenza:2024ehj} for reviews. Such axions are targets for a number of astrophysical observations ranging from CAST searches for axions from the Sun~\cite{CAST:2007jps,CAST:2017uph,CAST:2024eil} to X-ray spectral modulations~\cite{Marsh:2017yvc,Conlon:2017qcw,Reynolds:2019uqt,Reynes:2021bpe} to magnetic white dwarf polarization~\cite{Dessert:2022yqq} to X-ray searches for axion production and conversion in star clusters and galaxies~\cite{Dessert:2020lil,Ning:2024eky}, among many other possibilities~\cite{Meyer:2016wrm, Calore:2023srn, Lella:2024hfk, Meyer:2020vzy, Crnogorcevic:2021wyj, Hooper:2007bq}.  These axions, if they exist, must have $N = 0$ as otherwise their light masses would be fine-tuned (but see~\cite{Hook:2018jle}), though we note that this implies such ultralight axions with $E \neq 0$ are unlikely to arise in GUT theories~\cite{Agrawal:2022lsp,Agrawal:2024ejr}.

\textbf{\textit{Heavy axion induced gamma-ray flux from supernovae }} In this work we focus on core-collapse SN that lead to neutron stars (NSs).  The hot proto-NSs (PNSs) produce an abundance of axions, as described below, in the $\sim$10 s after collapse. After exiting the PNS photosphere, the axions decay into photons with the partial lifetime
\es{}{
\tau_{a\to\gamma\gamma} \approx 4\cdot 10^3 \, \, {\rm yr} \left( {100 \, \, {\rm MeV} \over m_a} \right)^3 \left( {10^{-15} \, \, {\rm GeV}^{-1} \over g_{a\gamma\gamma}}\right)^2 \,.
}
For $m_a \ge 3m_\pi$, with $m_\pi$ the pion mass, the decays $a\to 3\pi^0$ and $a\to \pi^0\pi^+\pi^-$ become kinematically possible. We account for these additional decay channels in our computation of the signal flux 
(see the SM for details).

For a SN at a distance $R_\mathrm{SN}$ from the Earth, the differential gamma-ray flux received on Earth (in units of {\it e.g.} ${\rm cts}/{\rm cm}^2/{\rm s}/{\rm GeV}$) is~\cite{Muller:2023vjm} 
\begin{align}
& \frac{\mathrm{d}^3 F_\gamma}{\mathrm{d} E_\gamma \mathrm{d} t \mathrm{~d} c_\theta}=\frac{2 }{\tau_{a\to\gamma\gamma}} \frac{\left|c_\theta\right|}{\left(t / R_{\mathrm{SN}}+1-c_\theta\right)^2} \frac{\mathrm{d} N_a / \mathrm{~d} E_a}{4 \pi R_{\mathrm{SN}}^2} \notag \\
& \times \frac{m_a}{p_a} \exp \left[-\frac{R_{\mathrm{SN}}}{\tau_a} \frac{2 E_\gamma}{m_a}\left(\frac{t}{R_{\mathrm{SN}}}+1-c_\theta\right)\right] \notag\\
& \times \Theta_{\text {cons. }}\left(E_\gamma, t, c_\theta\right) \,,
\label{eq:differential_flux}
\end{align}
where $E_\gamma$ is the observed gamma-ray energy, $t$ is the time delay relative to the arrival of a photon emitted at the onset of the SN explosion (in the limit $t$ much larger than the duration of the explosion itself, which is approximated as instantaneous), and $c_\theta \equiv \cos \theta$ with $\theta$ the angle between the gamma-ray arrival direction and the line-of-sight to the SN. 
We denote the axion momentum $p_a=\sqrt{E_a^2 - m_a^2}$ and total axion lifetime $\tau_a = (\Gamma_{a\to\gamma\gamma}+\Gamma_{a3\pi})^{-1}$.
Note that the flux is proportional to the axion emission spectrum $dN_a/dE_a$, where the axion energy $E_a$ is implicitly a function of $(E_\gamma, t, c_\theta)$. Above, we  also introduce a Heaviside function $\Theta_{\rm cons.}(E_\gamma,t,c_\theta)$ that enforces that the axion decays outside of the SN photosphere.  We emphasize that even light, relativistic axions can give time-delayed signals because the axions can leave the SN at large angles to the line of sight and then decay to photons that arrive at Earth with significant time delay. 

To compute the axion emission spectrum $dN_a / dE_a$, we follow Ref. \cite{Manzari:2024jns}. We use three different 1-D spherically symmetric SN simulations from the Garching Core Collapse Supernova Archive~\cite{GCCSN,Bollig:2020xdr}: SFHo-18.6, SFHo-18.8 and SFHo-20.0. Our fiducial model for SN1987A is, as in~\cite{Manzari:2024jns}, SFHo-18.6, which assumes an 18.6 $M_\odot$ progenitor and leads to a NS mass $1.553 M_\odot$, consistent with the inferred remnant mass for SN1987A~\cite{Page:2020gsx}.   
On the other hand, SN2023ixf and SN2024ggi are expected to have arisen from red supergiant (RSG) progenitor stars, unlike the case of SN1987A for which the SN was a blue supergiant (BSG), with masses $\sim$9 -- 14 $M_\odot$~\cite{Hosseinzadeh:2023ixa,2023ATel16050....1S, Pledger:2023ick,Kilpatrick:2023pse,Xiang:2024nee} (see SM Tab.~\ref{tab:properties} for a list of the known properties).  The low progenitor masses of these core collapse SN strongly suggest that the remnants are NSs and not black holes, though little information is known about the resulting NS masses. To be conservative we thus use the lowest NS progenitor masses from the suite of simulations we consider from the Garching archive; in particular, we use the SFHo-18.8 model for SN2023ixf and SN2024ggi, which has a $1.351M_\odot$ NS mass. In the SM we show that our results are not strongly affected by different choices of the SN simulation.   

We compute the axion luminosity integrated over the first $\sim$10 s of the SN as in~\cite{Manzari:2024jns} from the 1-D radial profiles, as shown in SM Fig. \ref{fig:SN_spectra} for two example axion masses and assuming the GUT benchmark scenario $E/N = 8/3$.
Following the discussion in Ref.~\cite{Manzari:2024jns}, we consider axion emission from the Primakoff process, 
bremsstrahlung from nucleons, and pion-to-axion conversion off nucleons, through a four-point interaction or via intermediate nucleon or $\Delta$ exchanges~\cite{Raffelt:1985nk, Carenza:2019pxu, Ho:2022oaw}. Additionally, we account also for photon-photon coalescence, which becomes kinematically allowed and relevant for $m_a \gtrsim 2\omega_{\rm pl}$, where $\omega_{\rm pl}$ denotes the photon plasma frequency. Note that the hadronic production processes dominate by multiple orders of magnitude over the production processes solely involving photons.
The axion luminosity could be further enhanced by density-dependent effects beyond what we account for~\cite{Springmann:2024mjp,Springmann:2024ret}.
Lastly, following Ref.~\cite{Lella:2023bfb}, we also account for the reabsorption of axions within the PNS.

\textbf{\textit{Fermi data analyses}} We analyze Fermi-LAT data collected since its launch in 2008 for evidence of heavy axion induced gamma-ray signals due to axion decay to two photons for the four SN targets considered in this work. Our analyses of Cas A and SN1987A are qualitatively different from those of SN2023ixf and SN2024ggi because the former SN occurred prior to the launch of Fermi while the latter two SN happened while Fermi was actively collecting data.  We begin by describing our analysis of the data for SN1987A. 

The axion lifetimes of primary interest for our study of SN1987A are much longer than the $\sim$35 years since the explosion of the SN. Thus, we stack the Fermi data -- subject to the binning described below -- over time for the period between 2008-08-04 and 2024-12-02. (Our data reduction procedure and set of quality cuts are fully described in App.~\ref{sec:data_reduction}.) We reduce and analyze the data for the top quartile of \texttt{SOURCE} data as ranked by the point spread function (PSF); we chose to not analyze the other quartiles of data for SN1987A because of their poorer angular resolution at low energies.  We bin the data spatially using \texttt{HEALPIX}~\cite{Gorski:2004by,Zonca:2019vzt} with ${\rm \texttt{nside}}=512$ and in energy using 10 logarithmically spaced energy bins between $100$ MeV and 1 GeV. 

\begin{figure}[!htb]
    \centering
    \includegraphics[width=0.48\textwidth]{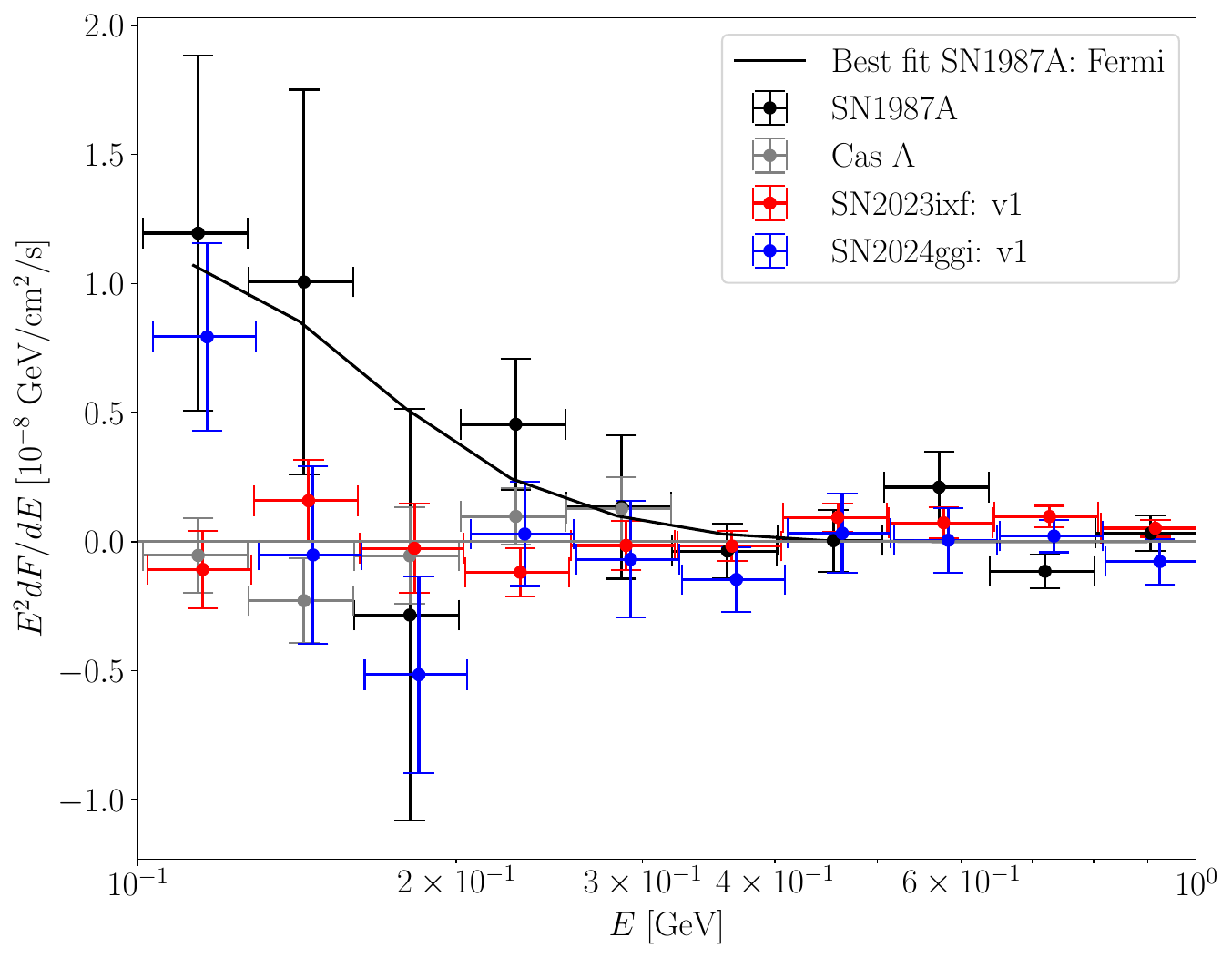}
    \caption{ The spectra from each of the targets considered in this work from analyses of the Fermi-LAT data since the launch of Fermi in 2008. For SN1987A and Cas A (which only includes data to $\sim$320 MeV due to astrophysical contamination at higher energies and is considered as a PS here) we stack in time all data since the SN, while for the two more recent SN we use the data prior to the SN as data-driven background templates to analyze the data post SN. Here, we present results for our \texttt{v1} data sets for SN2023ixf and SN2024ggi, which are the data sets containing data from the SN explosions until Dec. 2, 2024, though we also analyze shorter-term data sets to look for signals with shorter lifetimes (see text). We illustrate the best-fit axion-induced signal for SN1987A, which has $\sim$2$\sigma$ detection significance.  The best-fit axion parameters are $(m_a, g_{a\gamma\gamma}) \approx (234\, \, {\rm MeV}, 5.3 \times 10^{-15} \, \, {\rm GeV}^{-1})$ for $E/N = 8/3$. Data points are staggered in energy for presentation only.
    }
    \label{fig:Fermi_SN1987A_data}
\end{figure}

We analyze the data in each energy bin independently using a spatial template fit including all pixels in the region of interest (ROI) that have pixel centers within $5^\circ$ of SN1987A. We model SN1987A as a point source (PS) broadened only by the PSF, and we constrain emission associated with this source profiling over nuisance parameters for our background model. The background model includes a spatial template for Galactic diffuse gamma-ray emission, which we model using the Fermi \texttt{gll\_iem\_v07} (\texttt{p8r3}) diffuse model.  We account for unresolved extragalactic and instrumental backgrounds using an isotropic template that follows the exposure map. Additionally, we include all 6 of the PSs in the 4FGL PS catalog with greater than 3$\sigma$ detection significance~\cite{Fermi-LAT:2019yla,Ballet:2023qzs} within $5^\circ$ of the SN.  We also include the four extended emission templates within 5$^\circ$ of the SN.  
Each of these spatial background templates (12 in total) is assigned an independent nuisance parameter in each energy bin that rescales the expected flux associated with that template. We profile over the nuisance parameters associated with the nuisance templates in each energy bin, using a Poisson likelihood, to construct the profile likelihood for the flux associated with the signal template. (See App.~\ref{sec:data_reduction}.) The resulting best-fit fluxes and associated 1$\sigma$ uncertainties for SN1987A PS emission are shown in Fig.~\ref{fig:Fermi_SN1987A_data}.  We note that the lowest two energy bins have positive, central flux estimates that are 1.7$\sigma$ and $1.3$$\sigma$ away from zero, though given the low significance we are unable to claim a detection of emission associated with the SN in any individual energy bin. Furthermore, we emphasize that due to the number of nearby astrophysical PSs and extended emission regions any detection would be difficult to definitively associate with the SN and not other emission regions within the LMC. 

Given the set of profile likelihoods in the individual energy bins we construct a joint profile likelihood $p({\bm d} | {\mathcal M}_{\rm axion}, {\bm \theta})$, with ${\bm d}$ the observed Fermi data counts vector, as a function of the signal model ${\mathcal M}_{\rm axion}$ parameter vector ${\bm \theta}$ by the product of the profile likelihoods in the individual energy bins evaluated at the predicted signal flux.  Here, ${\bm \theta}$ is the combination of axion model parameters. In the GUT benchmark model with $E/N = 8/3$ we may express all couplings in terms of $g_{a\gamma\gamma}$. In this case, we find the best-fit axion mass and coupling combination $(m_a,g_{a\gamma\gamma}) \approx (234 \, \, {\rm MeV},5.3 \times 10^{-15} \, \, {\rm GeV}^{-1})$, with a discovery test statistic (TS) of ${\rm TS} \approx 6.4$. Given that the axion model is a 2-parameter model, this corresponds to a $p$-value of $p \approx 0.04$ (around 2$\sigma$ evidence in favor of the signal model). In Fig.~\ref{fig:Fermi_SN1987A_data} we illustrate the axion model prediction for the best-fit mass and coupling on top of the flux data. Note that at the best-fit point the axion lifetime is only $\sim$$10$ yrs, such that incorporating timing information, which we do not do in this work, could provide additional insight into the axion origin of the signal.

We set 95\% one-sided upper limits on $g_{a\gamma\gamma}$ as a function of $m_a$, illustrated in Fig.~\ref{fig:gayy_constraints} by the lower bound on the excluded region. Note that this analysis is not sensitive to lower axion masses than shown due to the kinematic and geometric constraints imposed by the fact that the signal must be time delayed by $\sim$20 years (see the SM). 
The upper bound on the excluded region arises from the fact that at fixed $m_a$ as one increases $|g_{a\gamma\gamma}|$ the lifetime of the axion falls and eventually becomes much less than the $\sim$20 years between the SN and the launch of Fermi.  Precisely, the upper boundary of the excluded region for SN1987A is set by applying Wilks' theorem to the profile likelihood at fixed $m_a$ as a function of $|g_{a\gamma\gamma}|$ expanding about the local best-fit value at large $|g_{a\gamma\gamma}|$ to determine the one-sided lower bound on $|g_{a\gamma\gamma}|$. (Note that if no lower boundary exists for the exclusion region, we do not attempt to construct an upper boundary.)  

Our analysis of Fermi-LAT data towards Cas A is similar to that of SN1987A with a few key differences. First, while there are only three significant 4FGL PSs near Cas A, the SN is much closer to the Galactic plane (at $\ell \approx 111.74^\circ$, $b \approx -2.13^\circ$).  We thus use a reduced-size ROI of 3$^\circ$ in radius to be conservative in mitigating mismodeling due to Galactic diffuse emission along the plane.  We also account for the fact that, depending on the axion parameters, the spatial extension of the axion-induced signal could surpass the spatial extension of the PSF (see SM Fig.~\ref{fig:Cas_A_angular_spread}); indeed, we find that the 68\% containment radius of axion-induced emission could reach $\sim$8$^\circ$ due to the long baseline between the SN, which occurred around 1674, and today.  Given fixed axion model parameters we self-consistently compute the spatial morphology of the axion signal, convolve it with the PSF, and then search for evidence of this spatial component in the Fermi-LAT data.  We  use only the first five energy bins ($E \lesssim 320$ MeV), since non-thermal emission associated with the SN shock-wave has been detected above $\sim$500 MeV~\cite{Funk:2010cg}.  We find no evidence for an axion-induced signal, with the discovery TS being ${\rm TS} \approx 1.4$ for a two-sided test and ${\rm TS} = 0$ for a one-sided test, since the best-fit coupling is negative. The upper limit under the null hypothesis from the Cas A analysis is shown in Fig.~\ref{fig:gayy_constraints}, while Fig.~\ref{fig:Fermi_SN1987A_data} shows the spectral data for the analysis assuming no spatial extension of the signal beyond the PSF.

Our analyses of SN2023ixf and SN2024ggi differ from those of SN1987A and Cas A because the former SN went off while Fermi was collecting data. We thus use the data collected by Fermi prior to the SN as a data-driven background template.  We model the data collected since the SN, which we refer to as the signal data sets, by a combination of the data-driven background template and the signal, which is modeled as a PS.  We construct three different signal data sets differing in the time interval of incorporated data.  Our \texttt{v1} signal data set incorporates data from 1 day prior to the first optical signal of the SN (2023-05-18 for SN2023ixf and 2024-04-11 for SN2024ggi) to 2024-12-02.  We chose 1 day prior to the optical signal from the SN to ensure that the SN itself is captured within the signal data set since the optical signal is expected to lag the SN by 
around 0.6 days for SN2023ixf and 0.3 days for SN2024ggi \cite{Kozyreva:2024ksv, Pessi:2024kfq}.
Our \texttt{v2} (\texttt{v3}) data sets only includes data to 30 (3) days after the onset of the optical signals; these data sets are useful for searching for shorter-lived axion signals.  

Given that the analyses of the more recent SN are more straightforward than those of SN1987A and Cas A, we analyze independently the data from the top 3 quartiles of \texttt{SOURCE} data as ranked by the PSF and then construct a joint likelihood.   
To combine the results from the \texttt{v1}, \texttt{v2}, \texttt{v3} data sets we take the strongest limit at a given mass from that ensemble (see the SM for individual limits).  In Fig.~\ref{fig:Fermi_SN1987A_data} we illustrate our recovered fluxes for the analyses of the \texttt{v1} data sets joined over PSF quartiles. 

We find no evidence for axions in our analyses of any of the data sets for SN2023ixf and SN2024ggi (${\rm TS} < 5$ in all cases).  
We illustrate the excluded regions  
from these SN in Fig.~\ref{fig:gayy_constraints}. For presentation purposes, we construct joint likelihoods for the \texttt{v1}, \texttt{v2}, and \texttt{v3} data sets over both SN and present the joint upper limits in Fig.~\ref{fig:gayy_constraints}. (See the SM for results shown individually for each SN.)

To determine the upper boundaries of the excluded regions for the two recent SN we account for two physical effects. First, Fermi is on a $\sim$96.5 minute orbit and has an instantaneous field of view that covers $\sim$1/5 of the sky. Given that we do not know the precise times of the SN, to be conservative, we thus assume that the effective area of Fermi is precisely zero for the first 90 minutes following the SN. (Beyond the first 90 minutes we approximate the effective area as constant in time.) This means that short-lived axion signals, even if bright after the explosion, contribute an exponentially suppressed amount of flux to our signal model if the axion lifetime is less than 90 minutes. (In contrast, Ref.~\cite{Muller:2023pip} analyses Fermi-LAT data towards SN2023ixf with the additional data quality cut \texttt{ABS(ROCK ANGLE)<52}, which leads to a roughly three-day gap in exposure; our sensitivity to short-lived axions is stronger than that of~\cite{Muller:2023pip} for SN2023ixf in part because we do not include this quality cut.)   Secondly, we account for the fact that axions must decay outside of the progenitor star. 

We note that there are noticeable features in the upper limits shown in Fig.~\ref{fig:gayy_constraints} for $m_a$ near the pion mass $m_\pi$~\cite{Lella:2024dmx}.  The reason is that $f_a |g_{a\gamma\gamma}|$ formally diverges as $m_a \to m_\pi$, while $f_a |g_{a\gamma\gamma}| \to 0$ as $m_a \to m_{a, {\rm res}}$, with $m_{a,{\rm res}} \approx 0.82 m_\pi$ in the GUT scenario~\cite{Bauer:2021mvw}. 

In Fig.~\ref{fig:gayy_constraints} we also show our results from our reinterpretation of the SN1987A SMM data in the context of the GUT axion model (see the SM), though the excluded region is subdominant compared to those from Fermi except at high $g_{a\gamma\gamma}$. On the other hand, in the scenario where the heavy axion only couples to QED and not to QCD, which we argue in this Letter is not a well motivated UV scenario, we show in the SM that the analyses of SN2023ixf and SN2024ggi Fermi data still give leading constraints but that the Fermi data towards SN1987A and Cas A are not constraining because we are not able to probe long enough lifetimes in that case for the axion to give an appreciable signal after the time delay since the SN. Thus, the application of Fermi data towards SN1987A and Cas A requires the axion to couple to QCD in addition to QED.

Note that in Fig.~\ref{fig:gayy_constraints} we illustrate in gray constraints from existing searches for heavy axions, most of which use SN as probes (an exception being the decay of the irreducible axion background produced through freeze-in in the early universe~\cite{Langhoff:2022bij}).  As detailed in the SM, we recompute the existing SN bounds \cite{Lella:2023bfb, Lucente:2020whw, Caputo:2022mah} to account for production through nucleons and trapping in the theory with $E/N = 8/3$; we find that constraints for the decay of the diffuse SN axion background are especially relevant.  Note also in Fig.~\ref{fig:gayy_constraints} our constraints are formally not valid in the gray region labeled `SN1987A$\nu$', though this region is already excluded, because in this region of parameter space the back-reaction of the axion energy loss is important and not accounted for.  We do not show constraints from more model-dependent searches that require the axion to be DM or make other early-Universe assumptions~\cite{Kohri:2017ljt,Porras-Bedmar:2024uql,Balazs:2022tjl,Foster:2021ngm,Roach:2022lgo,Calore:2022pks,Depta:2020wmr}.

\textbf{\textit{Discussion}}
\label{sec:discussion}
In this work we produce leading constraints on heavy axions that couple to both QED and QCD, as generically expected for axions much more massive than the QCD axion.  
We show these axions may be produced within the first few seconds after a core collapse SN and then decay at later times, giving long-lasting gamma-ray signals that may be observed days to years to centuries after the SN with instruments such as Fermi. We search for evidence of axion-induced gamma-rays from four recent and nearby core-collapse SN: Cas A, SN1987A, SN2023ixf, and SN2024ggi. We find no significant evidence for axion-induced gamma-rays.

Our work could be improved in the future by further incorporating temporal information into the data analysis instead of simply working with time-stacked data sets. In the future our results will likely be improved by new SN.  Most optimistically, a future Galactic SN would provide transformative sensitivity to an axion signal, as we show in the SM. To best take advantage of such a signal, however, we must work towards future gamma-ray observatories beyond the Fermi-LAT having full-sky continuous coverage to ensure no time gaps between the SN explosion and the gamma-ray observations~\cite{Manzari:2024jns}.

\textbf{\textit{Acknowledgements}}
We thank Francesca Calore, Maxim Pospelov, and Nick Rodd for helpful conversations. We thank Andrea Caputo, Pierluca Carenza, Damiano Fiorillo, Kevin Langhoff, Alessandro Lella, Alessandro Mirizzi, Orion Ning and Georg Raffelt for comments on the manuscript. 
J.B., Y.P., G.P., and B.R.S. are supported in part by the DOE award DESC0025293.  B.R.S. acknowledges support from the Alfred P. Sloan Foundation. This research used resources of the National Energy Research
Scientific Computing Center (NERSC), a U.S. Department of Energy Office of Science User Facility located
at Lawrence Berkeley National Laboratory, operated under Contract No. DE-AC02-05CH11231 using NERSC
award HEP-ERCAP0023978.

\bibliography{Bibliography}

\clearpage

\appendix

\section{Properties of the supernovae studied in this work.}
\label{sec:properties_SN}
In this work we derive constraints on the heavy-axion parameter space from gamma-ray observations of Cas A, SN1987A, SN2023ixf and SN2024ggi. SN1987A was a type II SN in the Large Magellanic Cloud, at a distance of $\sim$$50$ kpc from Earth. The progenitor star was a blue supergiant (BSG), Sanduleak -69 202, with radius $R\simeq 45\pm15\, R_{\odot}$. Recent observations and simulations of SN explosions and the evolution of NSs point toward a progenitor mass of $(18-20)\, M_{\odot}$ and a NS mass of $(1.22-1.62)\, M_{\odot}$~\cite{Page:2020gsx}. Less is known for SN2023ixf and SN2024ggi. The first was a type II-L SN located in the Pinwheel Galaxy, at a distance of $\sim$$6.4$ Mpc from Earth and has a candidate red supergiant (RSG) progenitor of approximately 11 $M_{\odot}$. The second was a type II SN in NGC3621 at a distance of $\sim$$7.2$ Mpc from Earth and has a candidate RSG progenitor of approximately 13 $M_{\odot}$. The masses of the remnant NSs are not known, and we assume that both SN have given birth to NSs with lower masses than SN1987A. On the other hand Cassiopea A (Cas A) is the remnant of SN1680 and the youngest SN remnant observed in the Milky Way at a distance of $d\approx 3.4$ kpc \cite{Milisavljevic:2024mbg}. Analysis of its X-ray spectrum suggests a radius $R \in [8.3,\, 10.3]$ km and a mass $M_\mathrm{NS} \in [1.65,\, 2.01]\, M_{\odot}$~\cite{Heinke_2010, Bonanno:2013oua}. Note that for sufficiently large axion-photon couplings the signal flux depends on the radius of the SN photosphere $R_\mathrm{ps}$. In our computations we assume $R_\mathrm{ps}$ and  $R_\mathrm{SN}$ take the central values of the ranges listed in SM Tab. \ref{tab:properties}.   

In this work we use the spherically-symmetric SN simulations including PNS convection~\cite{Mirizzi:2015eza,Fiorillo:2023frv} from~\cite{Bollig:2020xdr} that may be found in the Garching Core-Collapse Supernova archive~\cite{GCCSN} (see~\cite{Rampp:2002bq,Janka:2012wk, Bollig:2017lki} for further details of the incorporated microphysics). 
For SN1987A we use SFHo-18.6 as our fiducial simulation as it assumes an 18.6 ${\rm M}_\odot$  progenitor and a NS mass of 1.553 ${\rm M}_\odot$. For SN2023ixf and SN2024ggi our fiducial simulation is model SFHo-18.8 which  assumes an 18.8 ${\rm M}_\odot$  progenitor, and a NS mass of 1.351 ${\rm M}_\odot$. Finally we use SFHo-20.0 as our fiducial simulation for Cas A, which assumes a 20.0 ${\rm M}_\odot$ progenitor and a NS mass of 1.947 ${\rm M}_\odot$.

\section{Fermi data reduction and analysis}
\label{sec:data_reduction}
We use \texttt{FermiTools} to reduce 853 weeks of Pass 8 \textit{Fermi} gamma-ray data collected from August 4th, 2008, to December 2nd, 2024, restricted to the SOURCE event class photon classification. We apply the recommended quality cuts of \texttt{DATA\_QUAL>0}, \texttt{LAT\_CONFIG==1}, and \texttt{zenithangle<90}. We produce counts and exposure maps binned into \texttt{HEALPIX} maps with \texttt{nside=512}. The data are binned in 40 logarithmically spaced bins between 100 MeV and 1 TeV.  In Supp. Figs.~\ref{fig:residual_data}, \ref{fig:data_1987A}, \ref{fig:Cas_A}  
we show images of the data within the vicinity of each of our targets.

As described in the main Letter, we analyze the data differently for SN1987A and Cas A versus the two more recent SN in 2023 and 2024, since for the latter we may use the collected data prior to the SN as data-driven background models.  We begin by providing additional details for the SN1987A analysis. For that analysis, we analyze the data summed over the full time duration given above. We search for emission associated with a PS at the location of SN1987A. In that sense, the SN1987A search is analogous to the typical PS searches performed with Fermi-LAT data. However, the analysis is complicated by the number of gamma-ray sources in the LMC.  Due to the astrophysical gamma-ray backgrounds in the LMC, we chose to only analyze the top quartile of \texttt{SOURCE} data as ranked by the point spread function (PSF) for the SN1987A analysis. (See Supp. Figs.~\ref{fig:fermi_PSF} for illustrations of the PSF containment radii for different PSF quartiles.) This is because the lower-resolution data quartiles have degraded angular resolutions, which complicates the low-energy analysis in the complicated regions around SN1987A. This problem is exacerbated by the fact that we care about low energies down to 100 MeV and the Fermi PSF degrades significantly at low energies, making it all the more difficult to differentiate emission from SN1987A from emission from nearby sources.

We analyze the data within each energy bin independently in a $5^\circ$ radius ROI around SN1987A with a Poisson likelihood. Specifically, in energy bin $i$ we compute the likelihood
\es{}{
p_i({\bm d} | {\mathcal M}, {\bm \theta}) = \prod_{j=1}^{\rm N_{\rm pix}} {{\mu_{i,j}({\bm \theta})}^{n_{i,j}} \over n_{i,j} !} e^{-\mu_{i,j}({\bm \theta})} \,, 
}
where ${\bm d} = \{n_{i,j} \}$ is the data set consisting of counts in energy bin $i$ and pixel $j$. The mean expected number of counts $\mu_{i,j}$ is a function of the model parameter vector ${\bm \theta}$, which for the signal plus background model ${\mathcal M}$ is 
\es{}{
\mu_{i,j}({\bm \theta}) = &A_{{\rm sig},i} T^{\rm sig}_{i,j} + A_{{\rm bkg},i} T^{\rm bkg}_{i,j} + A_{{\rm iso},i} T^{\rm iso}_{i,j} \\
&+ \sum_{k=1}^{N_{\rm PS}} A_{{\rm PS},i}^k T^{{\rm PS},k}_{i,j} + \sum_{k=1}^{N_{\rm ext}} A_{{\rm ext},i}^k T^{{\rm ext},k}_{i,j} \,. 
}
Here, $N_{\rm PS}$ ($N_{\rm ext}$) denotes the number of PSs (extended emission regions) considered in the analysis. The templates $T^{\rm sig}$ ($T^{\rm iso}$) ($T^{\rm bkg}$) ($T^{\rm PS}$) ($T^{\rm ext}$) denote the appropriate spatial templates, at fixed energy index $i$, for emission associated with the source (diffuse model) (isotropic emission) (PSs) (extended emission).  The template $T^{\rm bkg}$ is the (\texttt{p8r3}) diffuse model, appropriately reprocessed for our data selection criterion.  The isotropic template is taken to follow the exposure map, which is produced in the data reduction process and is approximately constant over our ROI. We select all PSs in the 4FGL catalog within $5^\circ$ of SN1987A that have discovery significance larger than 3$\sigma$ in the 4FGL-DR4 PS catalog, and we include all four Fermi extended emission templates associated with the LMC~\cite{Fermi-LAT:2019yla,Ballet:2023qzs}.  There are 6 PSs that pass that cuts described above, which we include in our analysis, while the extended emission templates are \texttt{LMC-FarWest}, \texttt{LMC-30DorWest}, \texttt{LMC-Galaxy}, and \texttt{LMC-North}. The closest of the 6 PSs to SN1987A is \texttt{4FGL J0537.8-6909}, being $\sim$$0.23^\circ$ away from the SN.  This particular source is identified with the pulsar wind nebula N 157B, which also is known to produce high-energy gamma-rays extending above a TeV~\cite{HESS:2015hiz}.  Note that we forward model the signal (PS) template along with the other PS and extended emission templates through the instrument response using the \texttt{FermiTools}.  In total, our model ${\mathcal M}$, in a given energy bin, has 12 nuisance parameters for the background templates along with 1 signal parameter $A_{{\rm sig},i}$ for the signal flux associated with the PS in that energy bin.

We construct a suite of 10 profile likelihoods as functions of the $\{ A_{{\rm sig},i} \}_{i=1}^{10}$ by maximizing the likelihoods at fixed $A_{{\rm sig},i}$ over the ensemble of nuisance parameters (see, {\it e.g.},~\cite{Safdi:2022xkm}).  We refer to these profiled likelihoods as $p_i({\bm d} | A_{{\rm sig},i})$.  We use these profile likelihoods to extract the best-fit spectra and associated uncertainties for the SN. We also use these profile likelihoods for our analyses in the context of the axion model. 

The axion model predicts signal flux in each energy bin. We thus construct the likelihood
\es{eq:likelihood}{
p\left({\bm d} | \{ g_{a\gamma\gamma}, m_a, {E \over N} \} \right) = \prod_{i = 1}^{10} p_i\left({\bm d} | A_{{\rm sig},i}\left(g_{a\gamma\gamma}, m_a,{E \over N}\right)\right) \,,
}
where the product is over the 10 energy bins included in this analysis. The axion model predicts a unique signal flux in each energy bin, $A_{{\rm sig},i}(g_{a\gamma\gamma}, m_a)$, as a function of the axion mass and axion-photon coupling, for a given fixed value of $E/N$. 
We use the likelihood in~\eqref{eq:likelihood} to search for evidence in favor of the axion model and to constrain the parameter space, following the discussion in the main text.  The discovery TS is defined as twice the log-likelihood difference between the best-fit signal and null models for best-fit $\hat g_{a \gamma\gamma} >0$, with ${\rm TS} = 0$ for $\hat g_{a \gamma\gamma} <0$ for a one-sided test (see, {\it e.g.},~\cite{Safdi:2022xkm}).    

Our analysis of Cas A is analogous to that of SN1987A except for three differences: (i) we use a reduced ROI of 3$^\circ$ radius since the source is close to the Galactic plane; (ii) we account for the spatial extension of the signal, in addition to the PSF, at a given $g_{a\gamma\gamma}$, $m_a$, and energy bin using~\eqref{eq:differential_flux}; and (iii) we only include the first 5 energy bins, which includes data between 100 MeV and $\sim$320 MeV, given that high-energy gamma-ray associated with the SN remnant of Cas A have been detected, starting at $\sim$500 MeV~\cite{Funk:2010cg} and consistent with arising from particle acceleration at the edge of the expanding shock-wave.

Our analyses of the SN2023ixf and SN2024ggi data lead to profile likelihoods $p_i({\bm d} | A_{{\rm sig},i} )$ in the individual energy bins, as in the case of SN1987A, and from that point the analysis proceeds as in the case of SN1987A. However, the construction of the profile likelihoods themselves in the individual energy bins is qualitatively different for the more recent SN, since in these cases we have images of the sky pre-SN. First, we down-bin the data to $\texttt{nside} = 64$ such that our background data set has a large ($\gg 10$) number of counts per bin. The background data set is defined as the data set constructed from Fermi data since launch to 2 days prior to the SN.  We refer to the background data set as ${\bm d}_{\rm bkg} = \{ n^{\rm bkg}_{i,j} \}$. We construct three distinct `signal data sets', which are defined as the stacked data collected 1 day prior to the onset of the optical signal from the SN to 3 days after (30 days after) (until the present day) for the three data set versions.  The precise times for these data sets are given in Tab.~\ref{tab:time_ranges_Fermi}.  We refer to the three distinct signal data sets as ${\bm d}_{\rm sig}^{v1}$, ${\bm d}_{\rm sig}^{v2}$, and ${\bm d}_{\rm sig}^{v3}$ for the data set extending to the present day, 30 days after the SN, and 3 days after, respectively. 
We analyze the three data sets and then, at a given mass $m_a$, take the strongest limits over the ensemble of three data sets. This accounts for the fact that longer-lived axion signal will be better searched for using {\it e.g.} the \texttt{v1} data set, while shorter lived data sets might be better searched for using {\it e.g.} the \texttt{v2} data set. 

For a given signal data set we compute the likelihoods in the individual energy bins by 
\es{eq:p_joint}{
p_i( \{ {\bm d}_{\rm sig}, {\bm d}_{\rm bkg} \} | {\mathcal M}, {\bm \theta} ) = &\prod_{j=1}^{\rm N_{\rm pix}} {{\mu_{i,j}({\bm \theta})}^{n_{i,j}^{\rm sig}} \over n_{i,j}^{\rm sig} !} e^{-\mu_{i,j}({\bm \theta})} \times \\
&{{A_{i,j}^{\rm bkg}}^{n_{i,j}^{\rm bkg}} \over n_{i,j}^{\rm bkg} !} e^{-A_{i,j}^{\rm bkg}} \,,
}
with 
\es{eq:mu}{
\mu_{i,j} ({\bm \theta}) = A_{{\rm sig},i} T^{\rm sig}_{i,j} + {\mathcal A}^{\rm bkg}_i  A_{i,j}^{\rm bkg} {{\mathcal E}^{\rm sig}_{i,j} \over {\mathcal E}^{\rm bkg}_{i,j}} \,.
}
Here, $T^{\rm sig}_{i,j}$ is the signal template, as in the case of SN1987A, with model parameter $A_{{\rm sig},i}$.  Each pixel is assigned a nuisance parameter $A_{i,j}^{\rm bkg}$; this nuisance parameter is constrained by the number of background counts in the background data sets in that pixel and contributes to the signal model prediction through the ratio of exposure maps ${\mathcal E}$ between the signal and background data reductions.  We allow for an overall nuisance parameter ${\mathcal A}_i^{\rm bkg}$ in~\eqref{eq:mu} that takes a constant value over the pixels in a given energy bin to account for uncertainties in the exposure maps.  Note that the signal and background model ${\mathcal M}$ has, in a given energy bin $i$, the signal model parameter $A_{{\rm sig},i}$ in addition to a single nuisance parameter per pixel $A_{i,j}^{\rm bkg}$, along with the nuisance parameter ${\mathcal A}_i$.

We may analytically profile over the nuisance parameters $A_{i,j}^{\rm bkg}$ appearing in~\eqref{eq:p_joint} to compute an analytic, though complicated, expression for the likelihood partially profiled over all nuisance parameters except for ${\mathcal A}_i^{\rm bkg}$. We profile over ${\mathcal A}_i^{\rm bkg}$ numerically in order to compute the profile likelihoods $p_i({\bm d} | A_{{\rm sig},i})$.

\clearpage

\unappendix

\clearpage
\onecolumngrid

\begin{center}
  \textbf{\large Supplementary Materials for ``Time-delayed gamma-ray signatures of heavy axions from core-collapse supernovae''}\\[.2cm]
  \vspace{0.05in}
  {Joshua N. Benabou, Claudio Andrea Manzari, Yujin Park, Garima Prabhakar, Benjamin R. Safdi, and Inbar Savoray}
\end{center}

These Supplementary Materials (SM) are organized as follows. Section~\ref{sec:supp_figures} contains supplementary figures and tables that support the results of the main text.  Section~\ref{sec:electron_decays} discusses additional axion decay channels beyond two photons.  In Sec.~\ref{sec:heavy_axion} we discuss aspects of heavy axion theory related to the strong-{\it CP} problem. In Sec.~\ref{sec:RGflow} we describe the massive axion couplings under RG flow, while Sec.~\ref{sec:absorption} provides additional details for our axion production and reabsorption calculations. Section~\ref{SM:additional_analyis_details} gives additional analysis details for the analyses presented in the main text.  Lastly, in Sec.~\ref{sec:other} we describe our computations of the other SN-related constraints shown in {\it e.g.} Fig.~\ref{fig:gayy_constraints}.  

\section{Supplementary Figures and Tables}
\label{sec:supp_figures}

\begin{figure}[!htb]
    \includegraphics[width=0.48\textwidth]{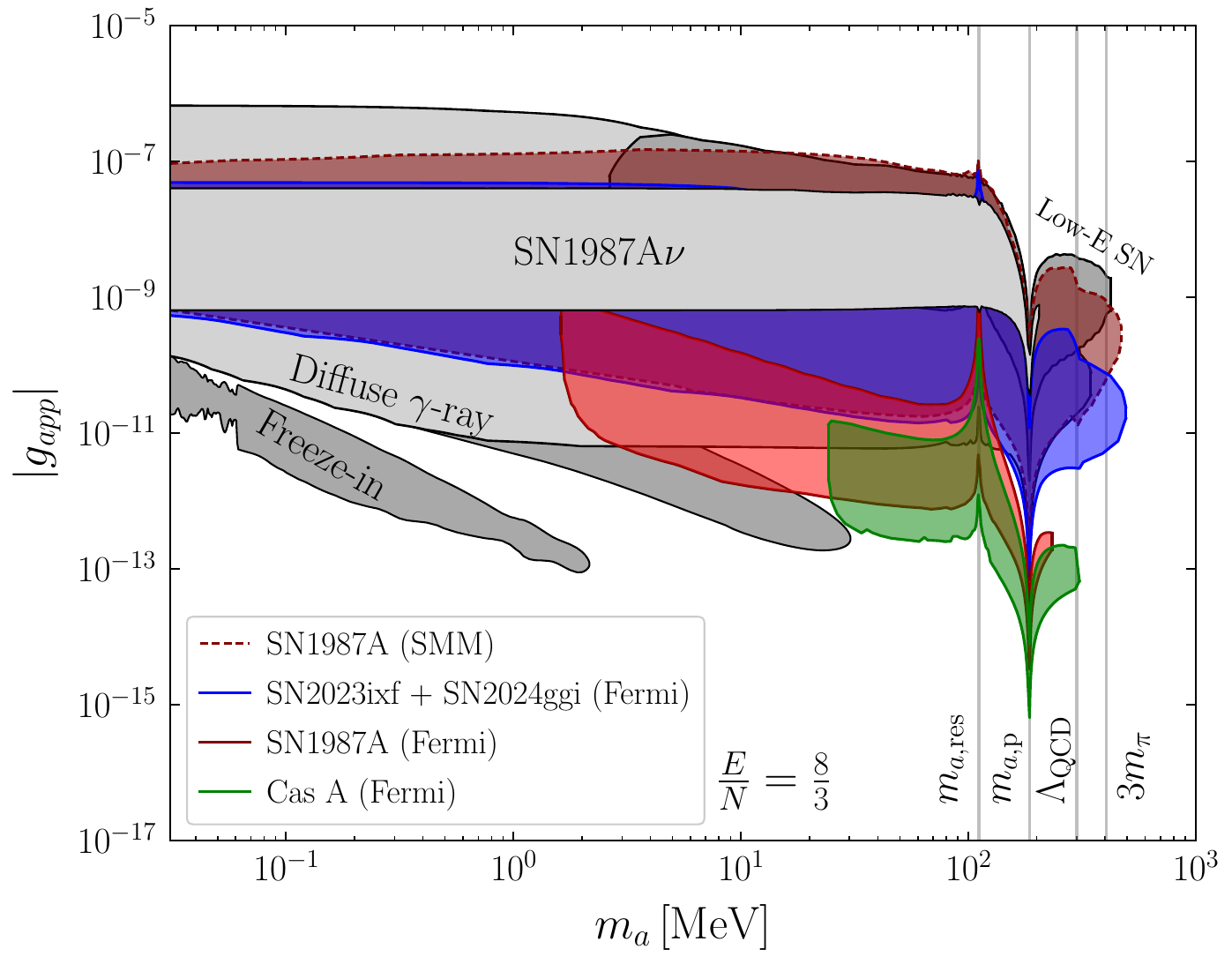}
     \includegraphics[width=0.48\textwidth]{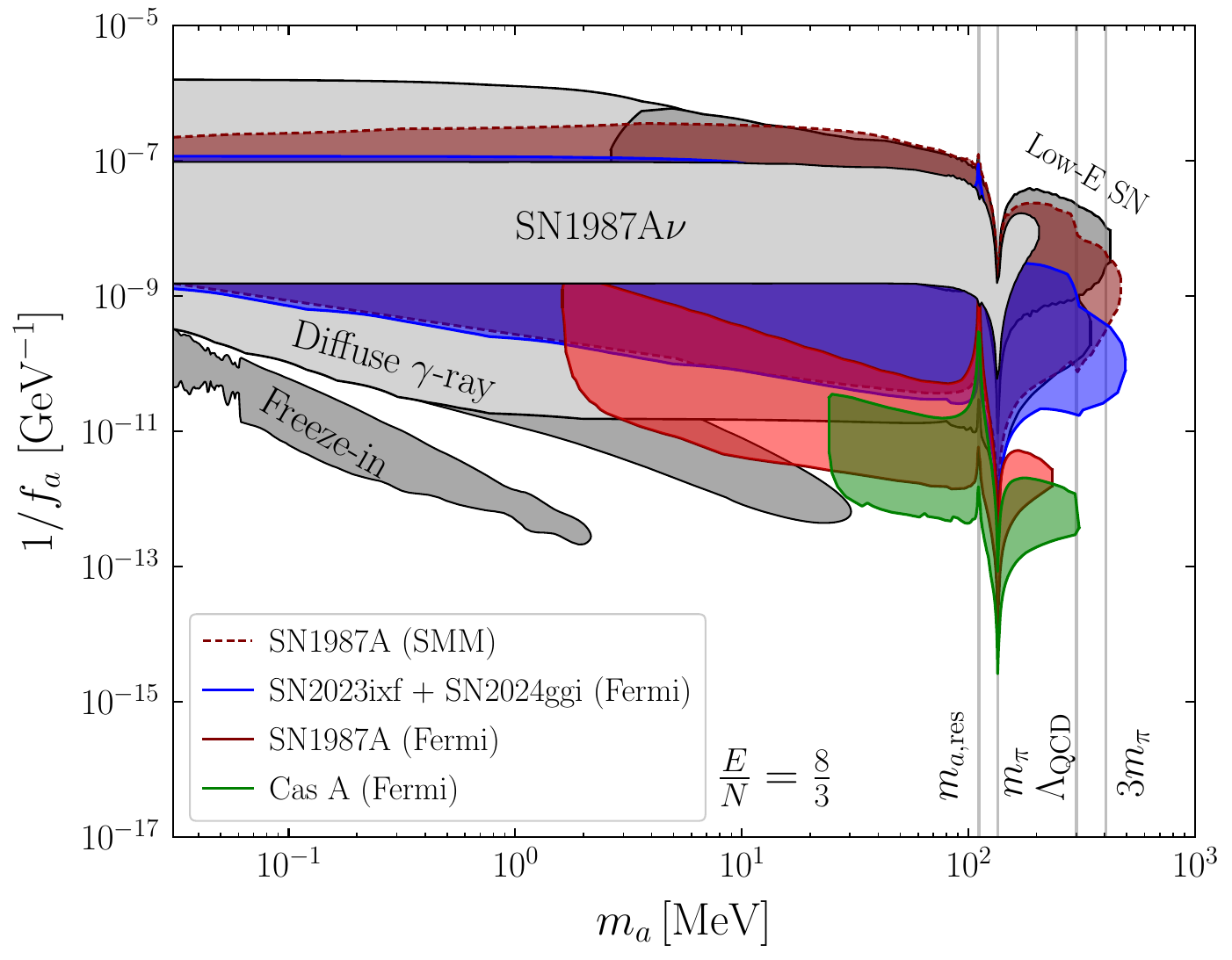}
        \caption{ 
        (\textit{Left}) Excluded regions found in this work (at 95\% confidence or greater) as in Fig.~\ref{fig:gayy_constraints} but in the $m_a$-$|g_{app}|$ plane. Here we define $g_{app}=C_{app}m_p/f_a$ with $m_p$ the proton mass. We indicate the axion mass $m_{a,\mathrm{p}}=185.9$ MeV for which $C_{app}$ is vanishing (see Fig.~\ref{fig:nucleon_couplings_GUT}).
        (\textit{Right}) As in the left panel, but in the $m_a$-$1/f_a$ plane. For an explanation of the regions of sharply enhanced or suppressed sensitivity, see Sec. \ref{sec:RGflow}.
    }
    \label{fig:gapp_constraints}
\end{figure}

\begin{figure}[h]
    \centering 
    \includegraphics[width=0.48\textwidth]{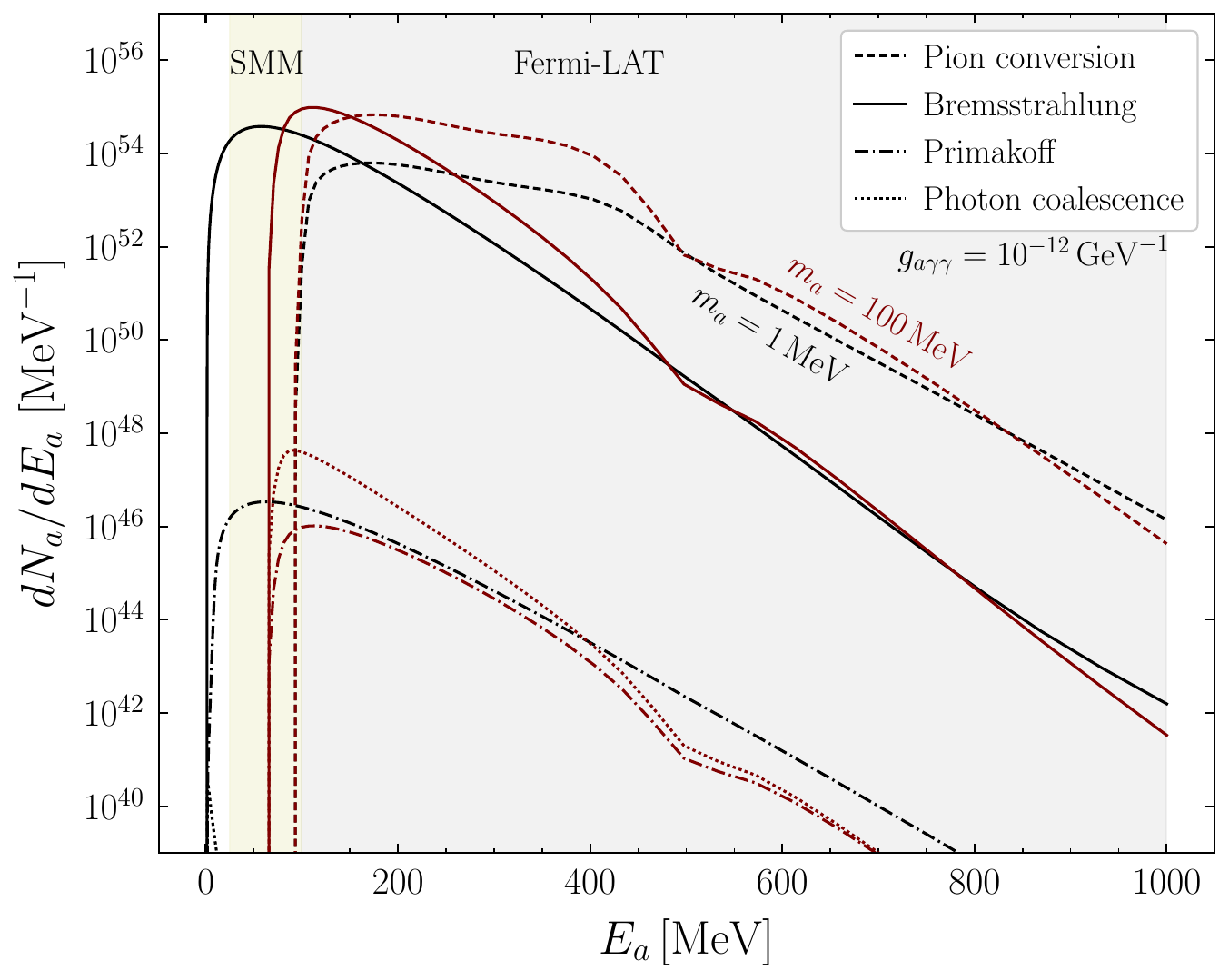}
    \caption{The axion emission spectra integrated over the first $\sim 10$ s of the SN, assuming the GUT benchmark model ($E/N = 8/3$), for various processes: photon coalescence, Primakoff production, nucleon-nucleon bremsstrahlung, and pion conversion. Here we fix $g_{a\gamma\gamma} = 10^{-12} \, \mathrm{GeV}^{-1}$, and show the spectra for $m_a = 1 \, \mathrm{MeV}$ (black) and $m_a = 100 \, \mathrm{MeV}$ (maroon). We compute SN properties using the SFHo-18.6 SN simulation. 
    The energy ranges we use in our analyses are indicated (shaded) for the SMM and Fermi-LAT observations.
    }
    \label{fig:SN_spectra}
\end{figure}

\begin{figure}[h]
    \centering
    \includegraphics[width=0.48\textwidth]{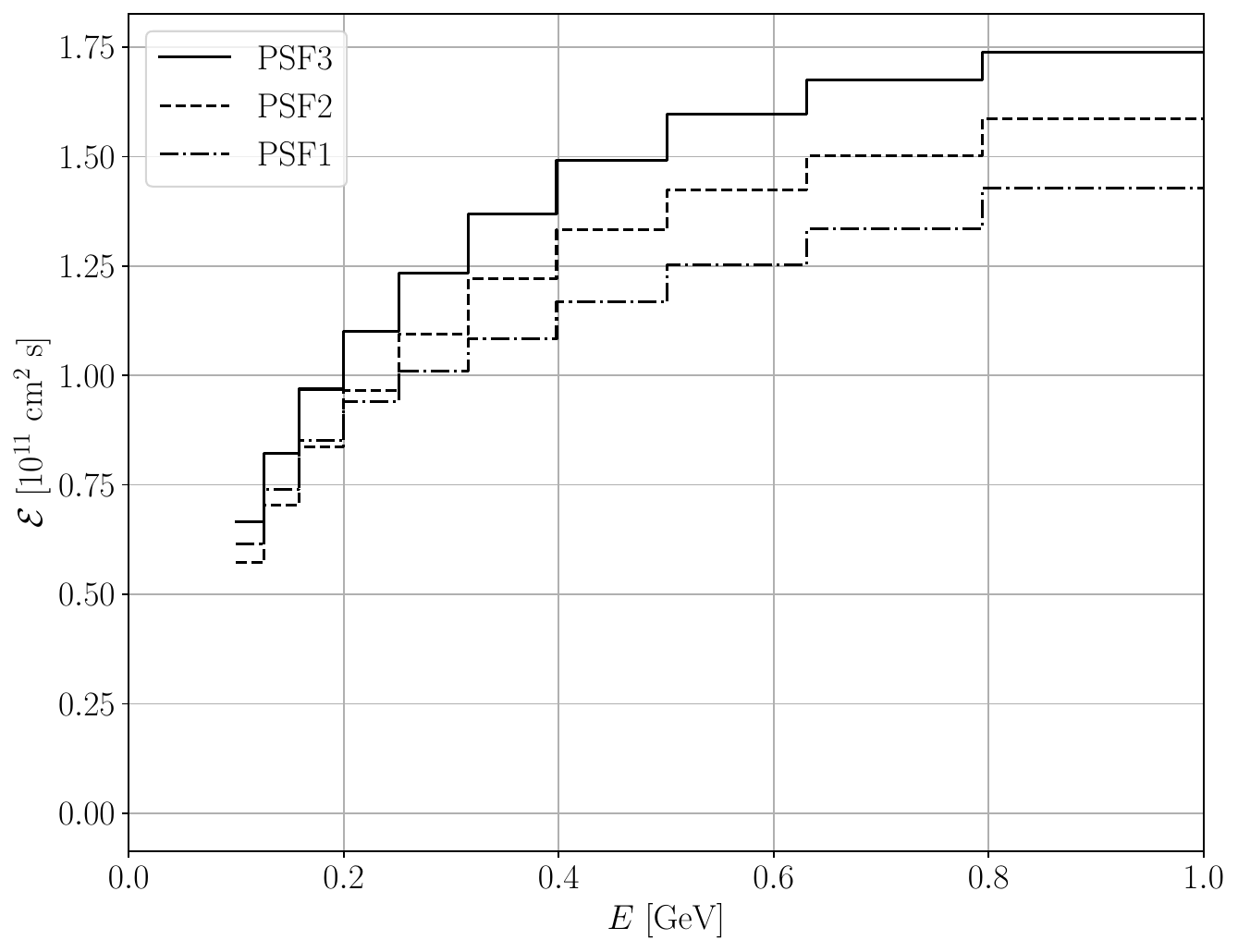}\includegraphics[width=0.48\textwidth]{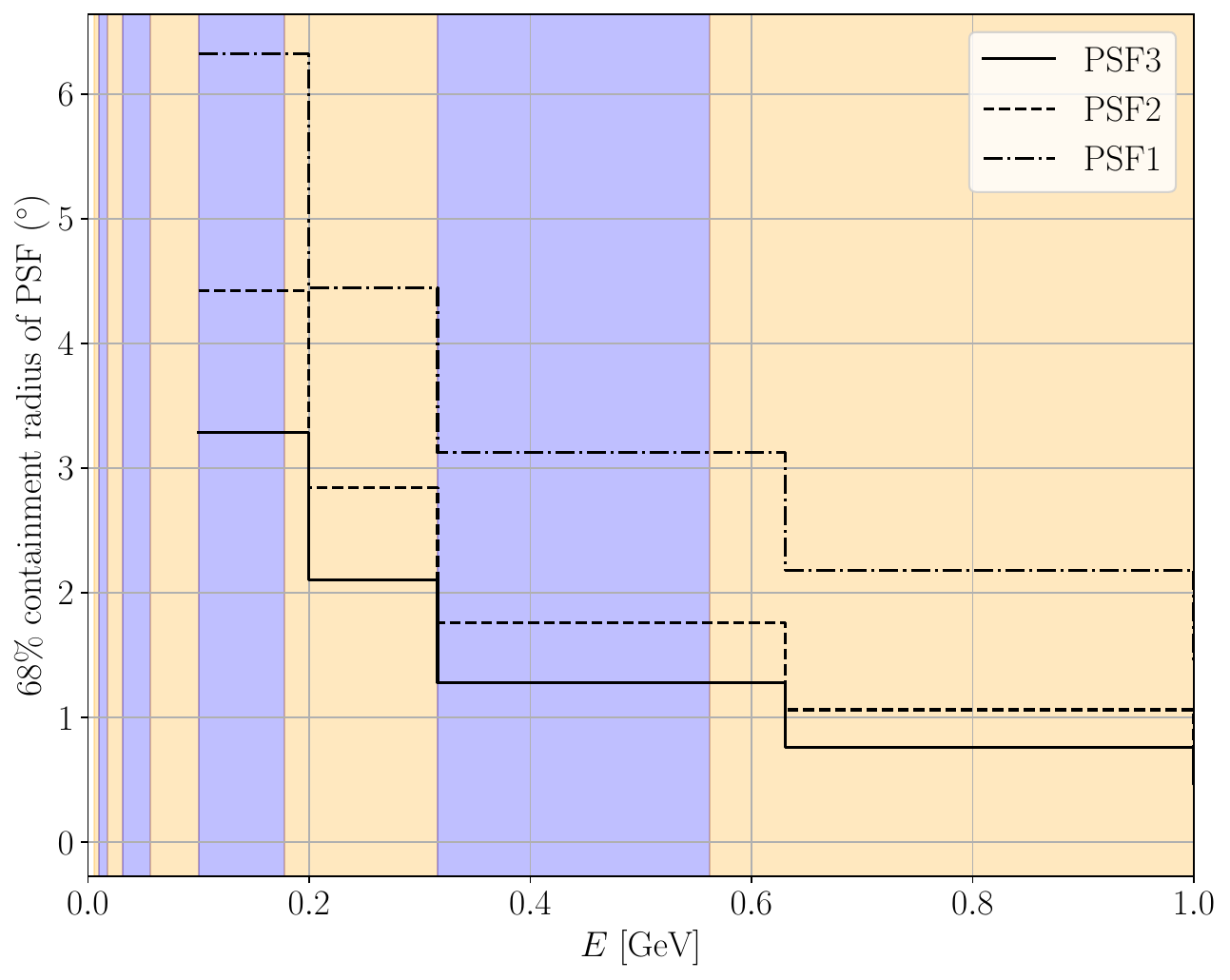}
    \caption{Aspects of the Fermi-LAT instrument response. (\textit{Left}) The exposure ${\mathcal E}$ in each of our analysis energy bins in units of cm$^2$s for each of the PSF quartiles, with PSF3 having the highest angular resolution and PSF1 the poorest resolution. (There is also a PSF0 quartile, but we do not include it since the angular resolution of this quartile is especially poor.) (\textit{Right}) The 68\% containment radius of the PSF in each of our analysis energy bins for the three PSF quartiles we consider. The Fermi PSF is modeled through King functions, but the \texttt{FermiTools} only includes the parameters of the King functions in coarse energy bins, which are shown alternating in orange and blue in the figure. Thus, within the orange and blue bands, we approximate the PSF as constant as a function of energy. 
    }
    \label{fig:fermi_PSF}
\end{figure}

\begin{figure}[h]
    \centering
    \includegraphics[width=0.48\textwidth]{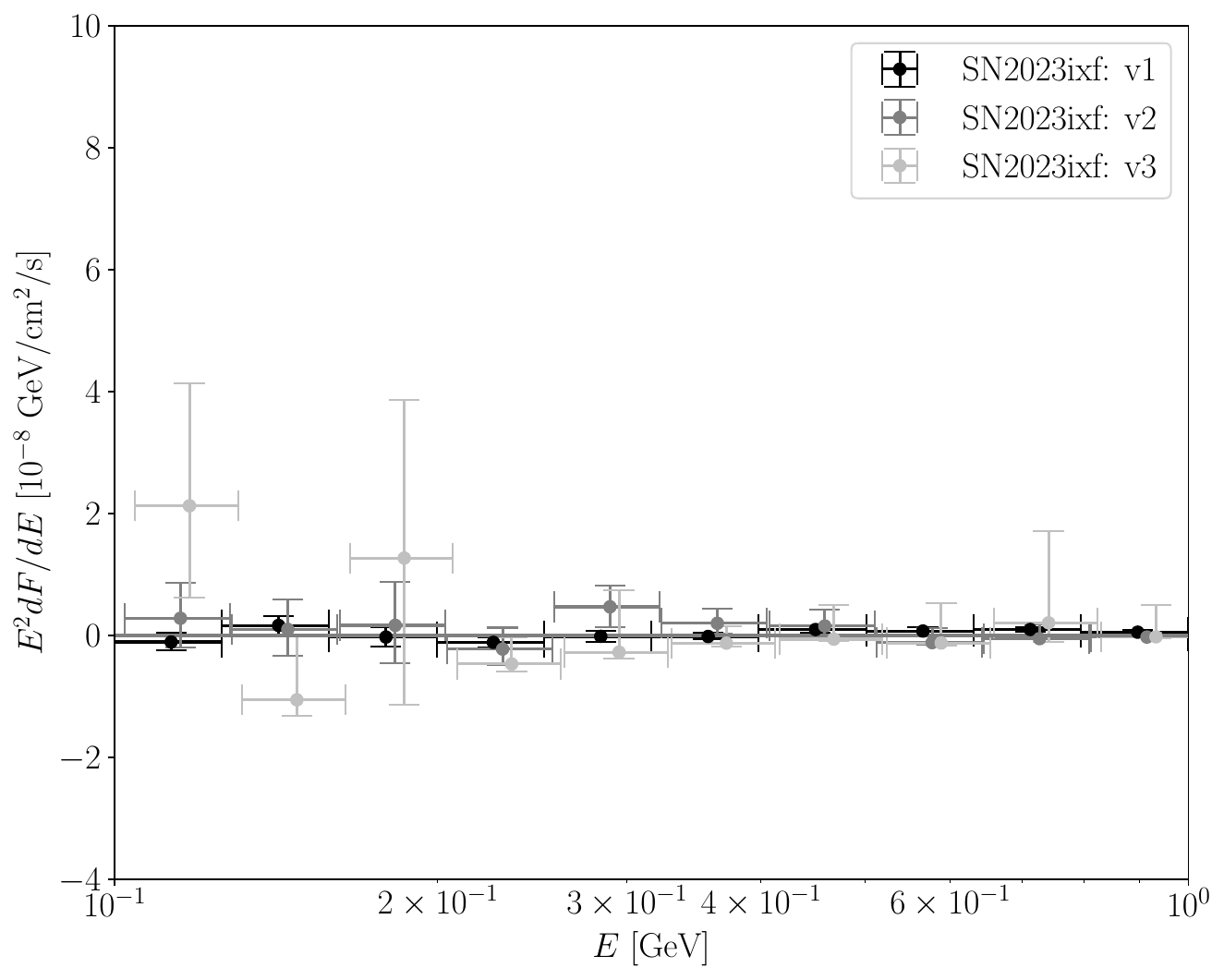}\includegraphics[width=0.48\textwidth]{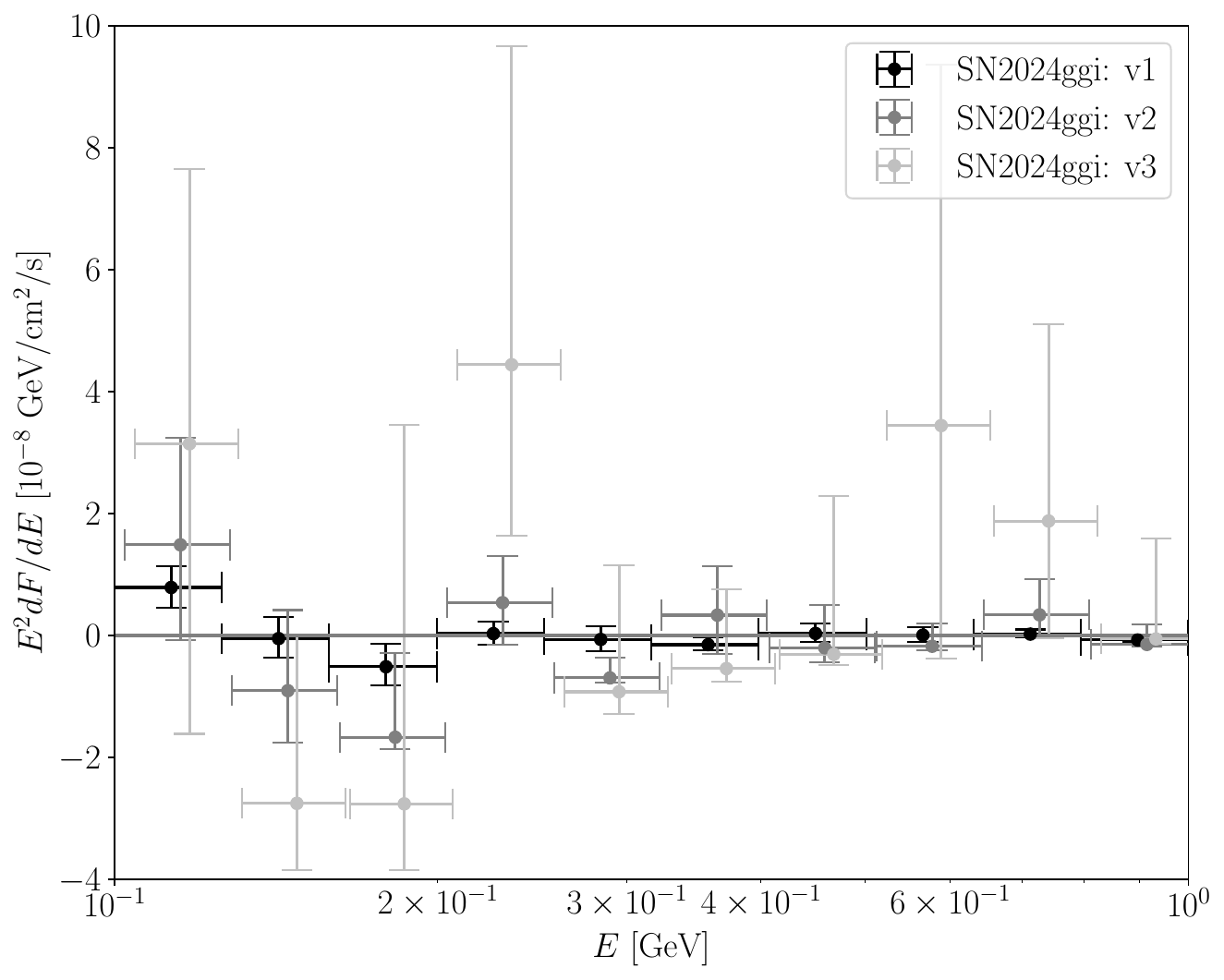}
    \caption{As in Fig.~\ref{fig:Fermi_SN1987A_data} except comparing the \texttt{v1}, \texttt{v2}, and \texttt{v3} data sets for SN2023ixf (\textit{Left}) and SN2024ggi (\textit{Right}). Recall that the \texttt{v1} data sets include data from slightly before the optical signal until December 2024, while \texttt{v2} (\texttt{v3}) extends to 30 days (3 days) past the SN.  Interestingly, SN2023ixf has smaller error bars than SN2024ggi even for the \texttt{v2} and \texttt{v3} data sets. This is due to the current observing strategy of the Fermi-LAT, which gives more exposure towards the location of SN2023ixf than SN2024ggi.
    }
    \label{fig:2023_2024_v123}
\end{figure}

\begin{figure}[h]
    \centering
    \includegraphics[width=0.48\textwidth]{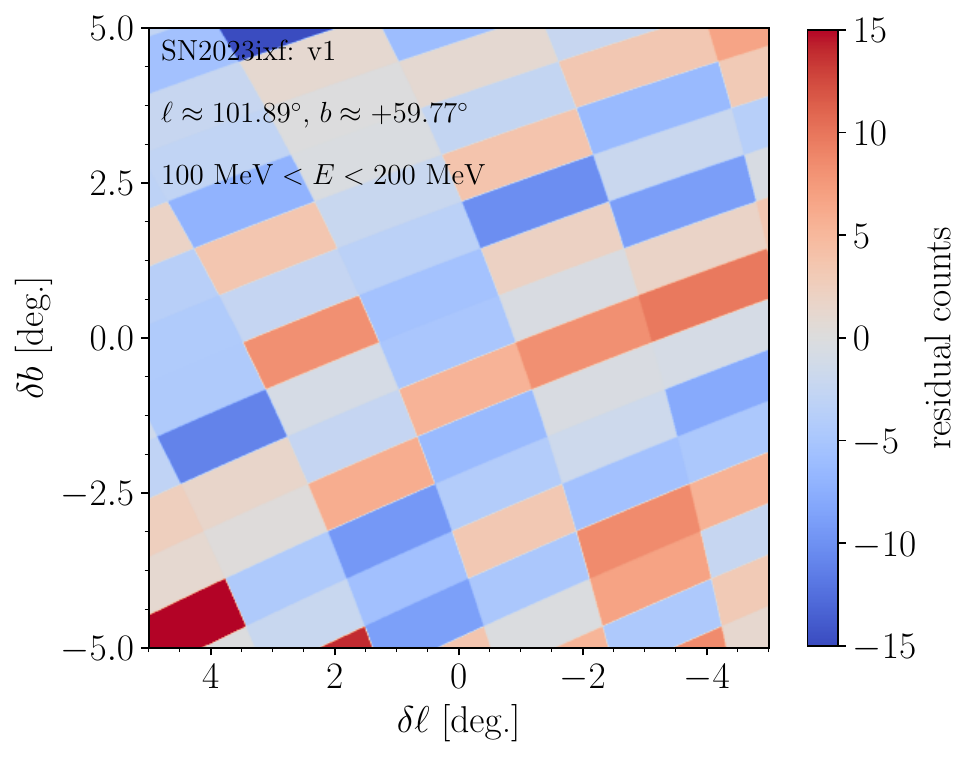}\includegraphics[width=0.48\textwidth]{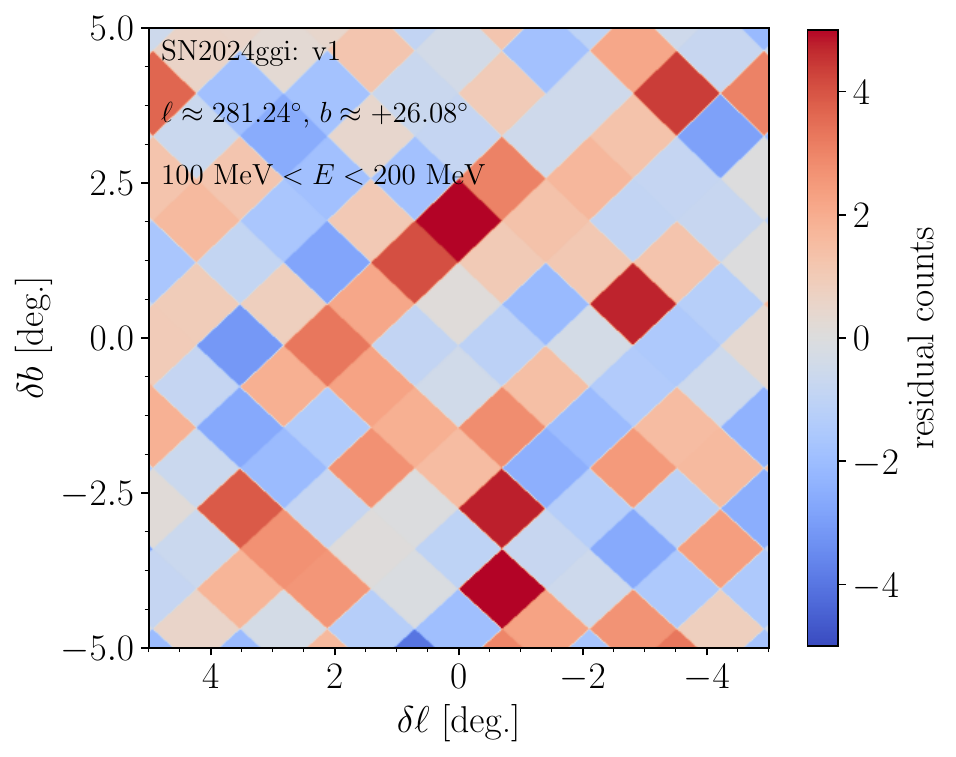}
    \caption{
    The SN2023ixf (\textit{Left}) and SN2024ggi (\textit{Right}) Fermi-LAT background-subtracted data between 100 and 200 MeV in the vicinity of the targets. Note that $\delta \ell$ and $\delta b$ refer to the difference in longitude and latitude from the source, respectively, with the source locations indicated in the figures. The background model arises from the data collected by the Fermi-LAT prior to the SN. We find no evidence for excess emission associated with either of these SN. Note that the data are down-binned here to $\texttt{nside}=64$ for presentation purposes.
    }
    \label{fig:residual_data}
\end{figure}

\begin{figure}[h]
    \centering
    \includegraphics[width=0.48\textwidth]{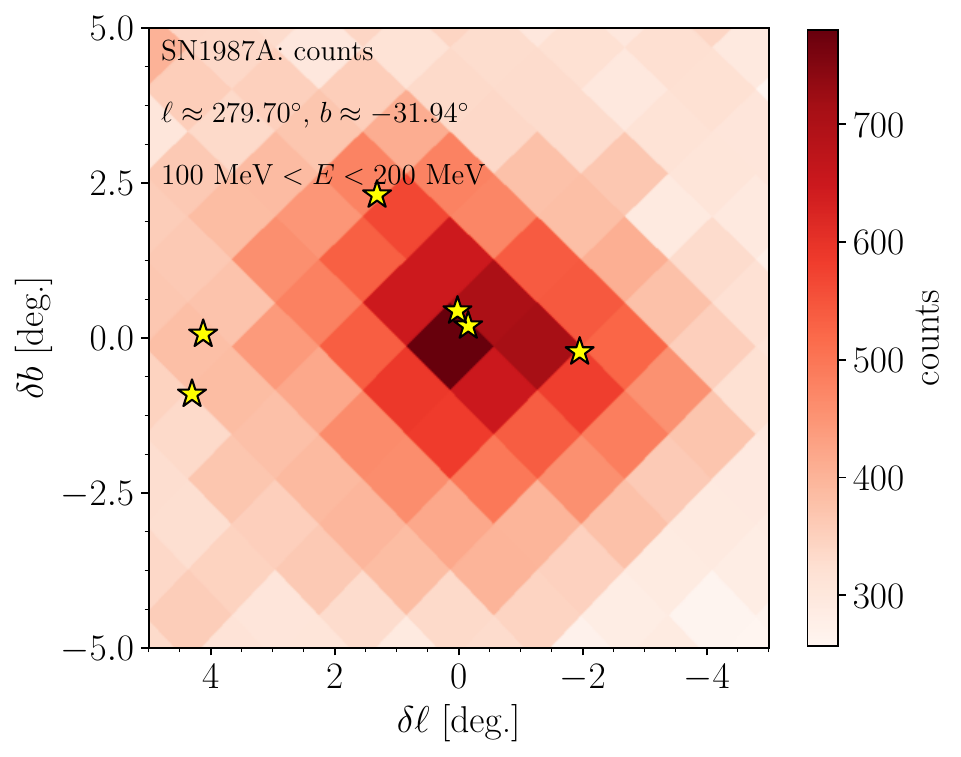}\includegraphics[width=0.48\textwidth]{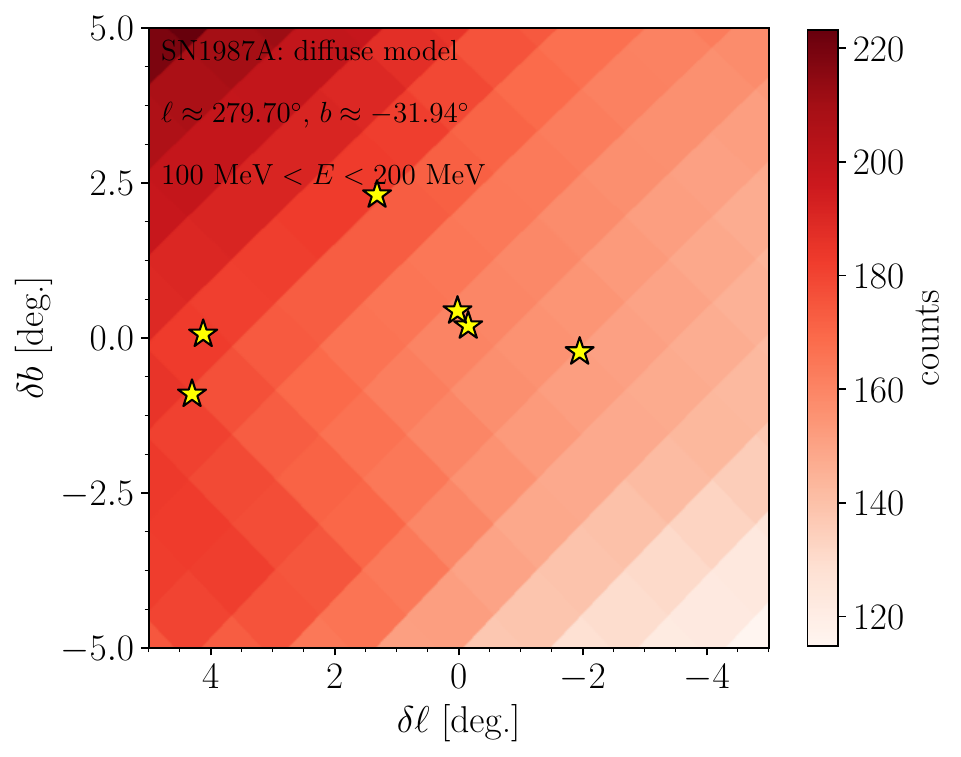}\\
    \includegraphics[width=0.4\textwidth]{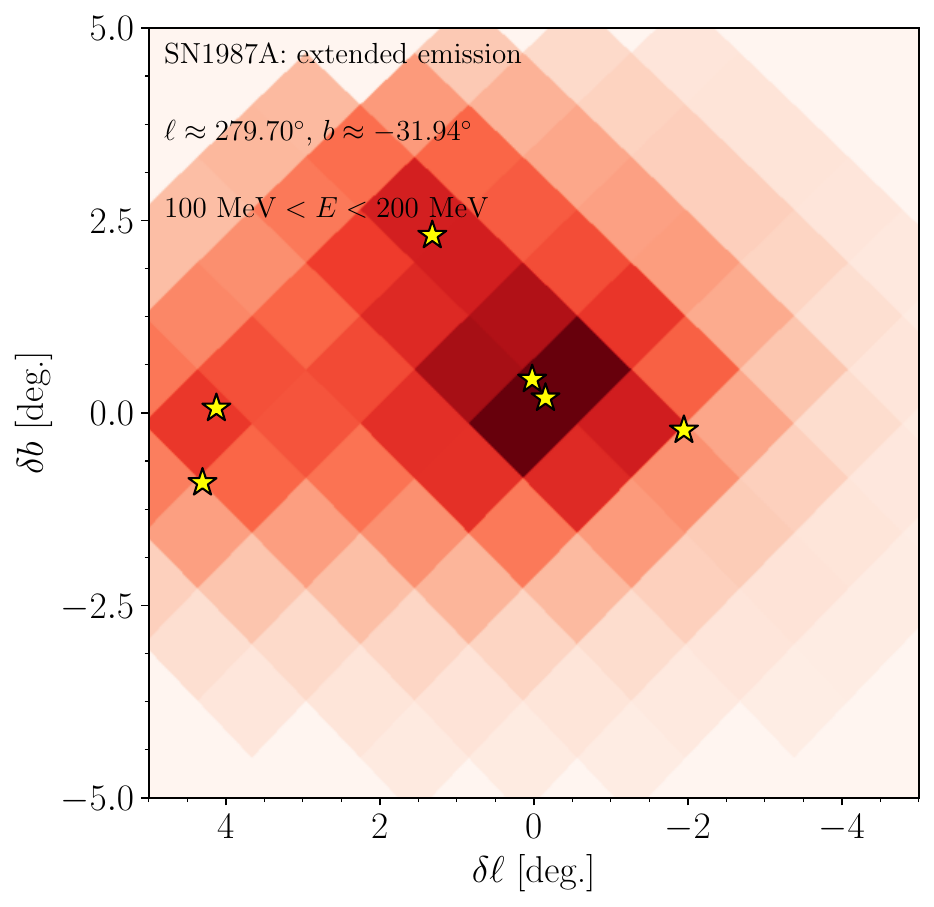}
    \caption{As in Fig.~\ref{fig:residual_data} but for the data in the vicinity of SN1987A. (\textit{Top left}) The counts between 100 and 200 MeV in the vicinity of SN1987A. The stars indicate the locations of the six astrophysical PSs we account for in our analysis. Note that two of these sources are especially close to the location of SN1987A. (\textit{Top right}) The Galactic diffuse model in the vicinity of SN1987A. (\textit{Bottom}) The sum of extended emission templates associated with the LMC with arbitrary normalization, though we note that in our analysis the normalization of each of the extended emission templates is an independent nuisance parameter.  As in Fig.~\ref{fig:residual_data}, we down-bin to $\texttt{nside}=64$ for presentation purposes.   
    }
    \label{fig:data_1987A}
\end{figure}

\begin{figure}[h]
    \centering
    \includegraphics[width=0.48\textwidth]{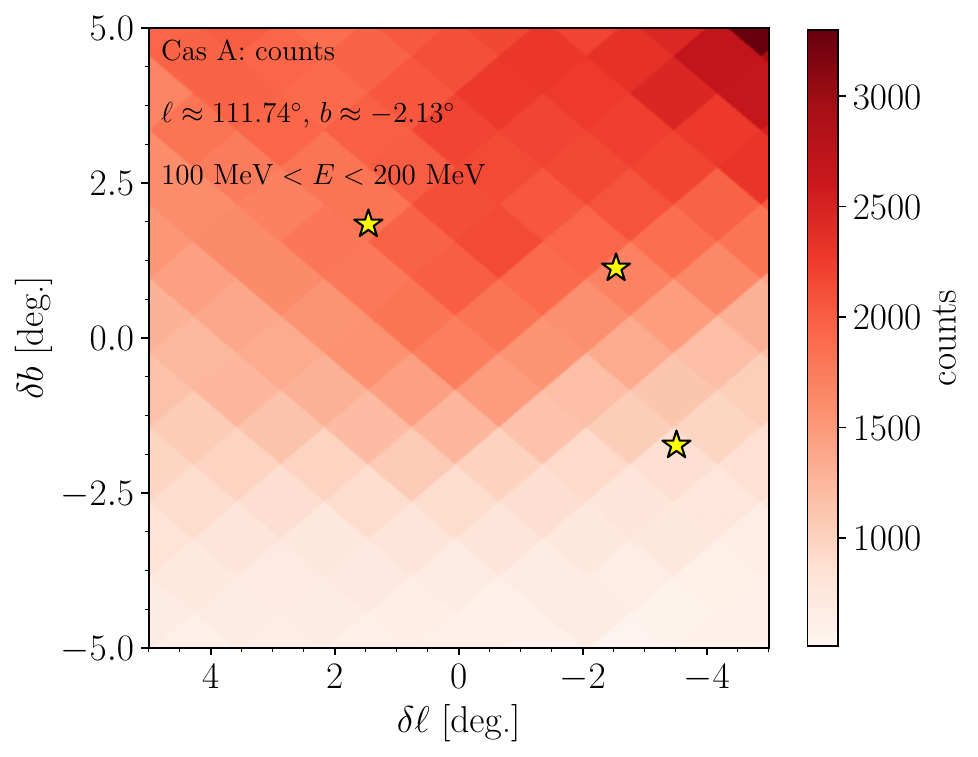}\includegraphics[width=0.48\textwidth]{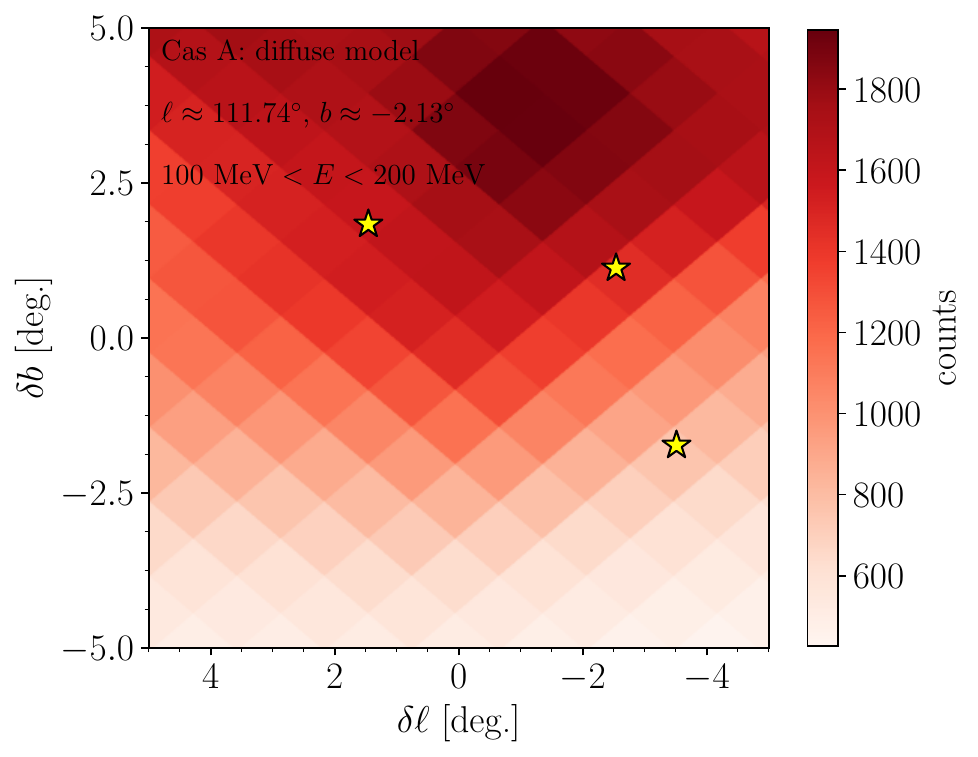}
    \caption{As in Fig.~\ref{fig:data_1987A} but for Cas A.  We have no extended emission templates for our Cas A analysis.
    }
    \label{fig:Cas_A}
\end{figure}

\begin{figure}[h]
    \centering
    \includegraphics[width=0.5\textwidth]{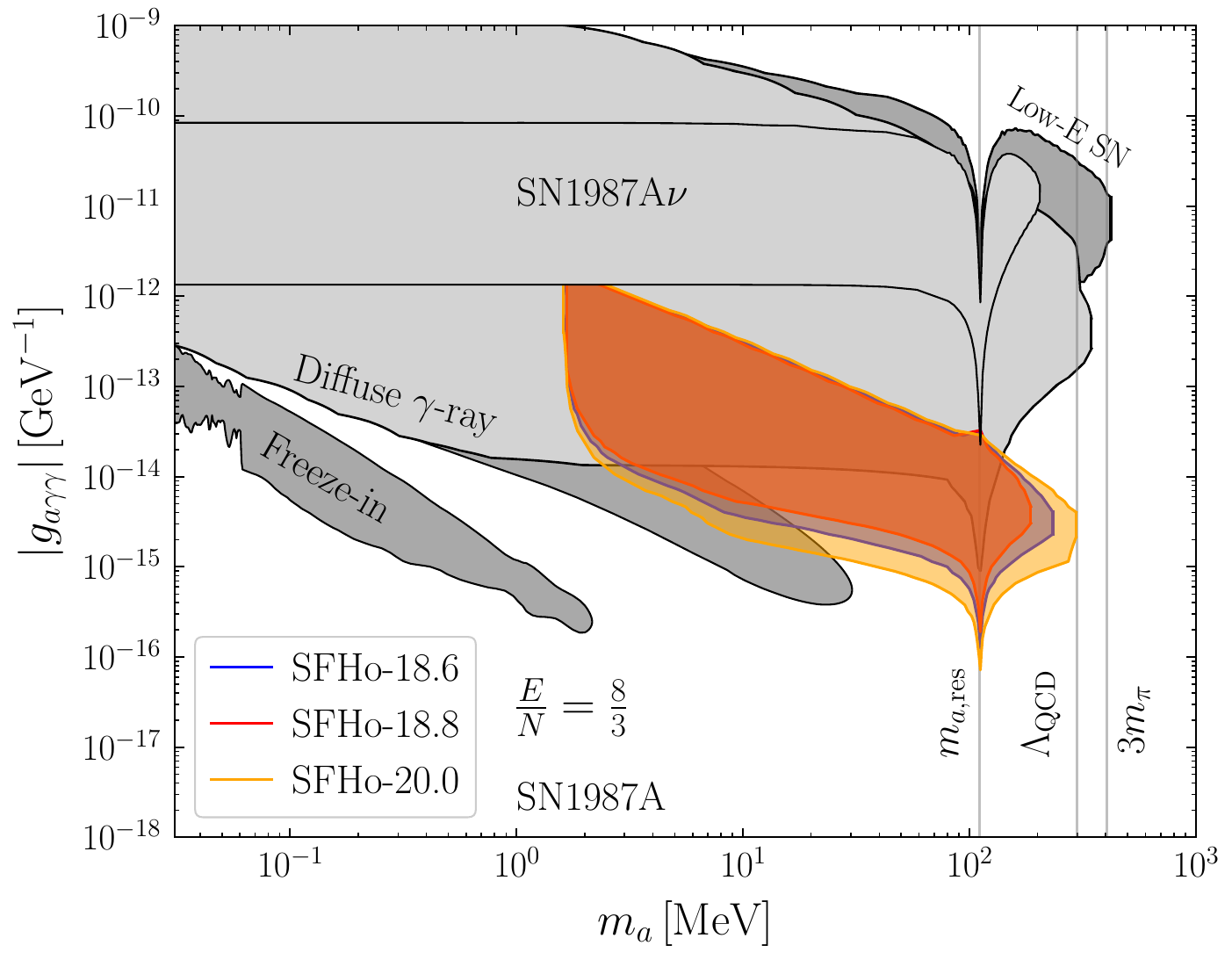}
    \includegraphics[width=0.48\textwidth]{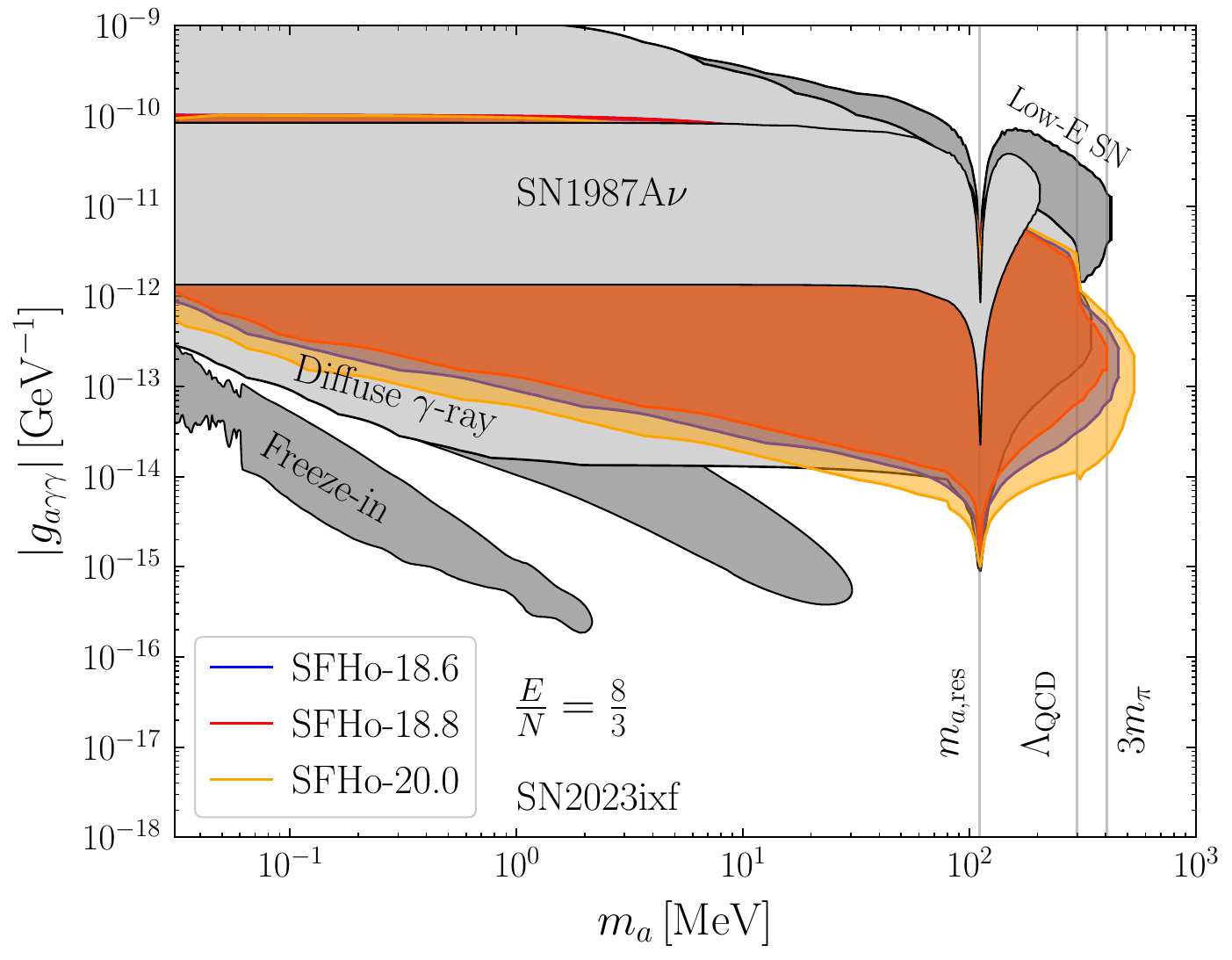}
    \caption{ 
    (\textit{Left}) The SN1987A results from this work computed with all analysis details fixed at their fiducial choices except for the choice of SN simulation. For details of the SN simulations see SM Sec. \ref{sec:properties_SN}. (\textit{Right}) As in the left panel but for SN2023ixf. 
    }
    \label{fig:varying_SN_sim_1987A}
\end{figure}

\begin{figure}[h]
    \centering
    \includegraphics[width=0.5\textwidth]{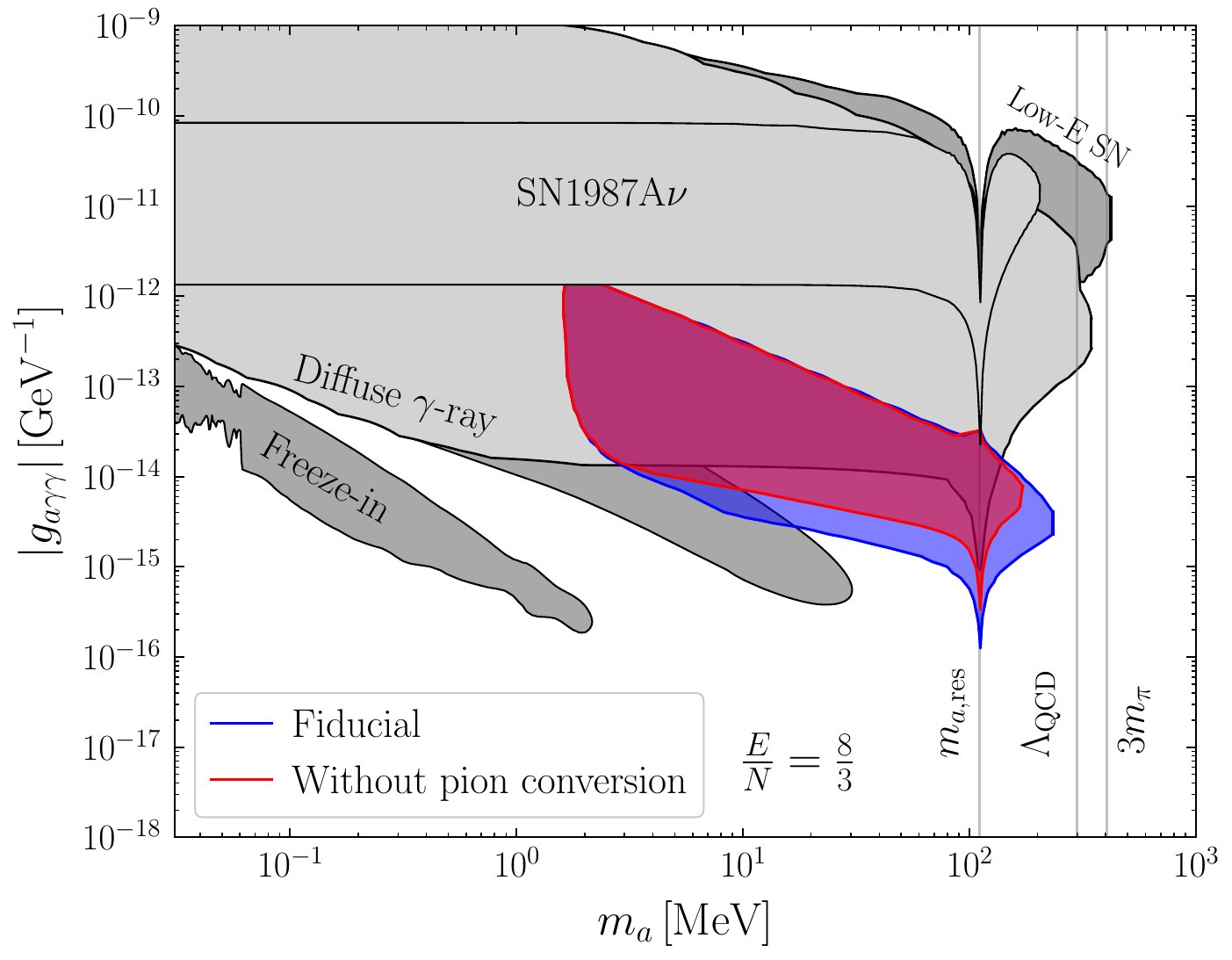}
    \caption{ 
    Our fiducial result from the Fermi-LAT analysis of SN1987A including all  processes relevant for axion production in the PNS (blue), compared to result obtained by not including any axion-pion conversion processes (red).  Pionic processes make an important contribution to our results, though even without accounting for pions we still find leading limits. We emphasize that pions are not included within the SN simulations we use; rather, we assume a $\pi^-$ population that is in weak equilibrium with the other particles in the PNS. Dedicated simulations incorporating pions are needed to further assess the impact of pions on the axion luminosity. 
    }
    \label{fig:comparison_pion_conversion}
\end{figure}

\begin{figure}[h]
    \centering
    \includegraphics[width=0.48\textwidth]{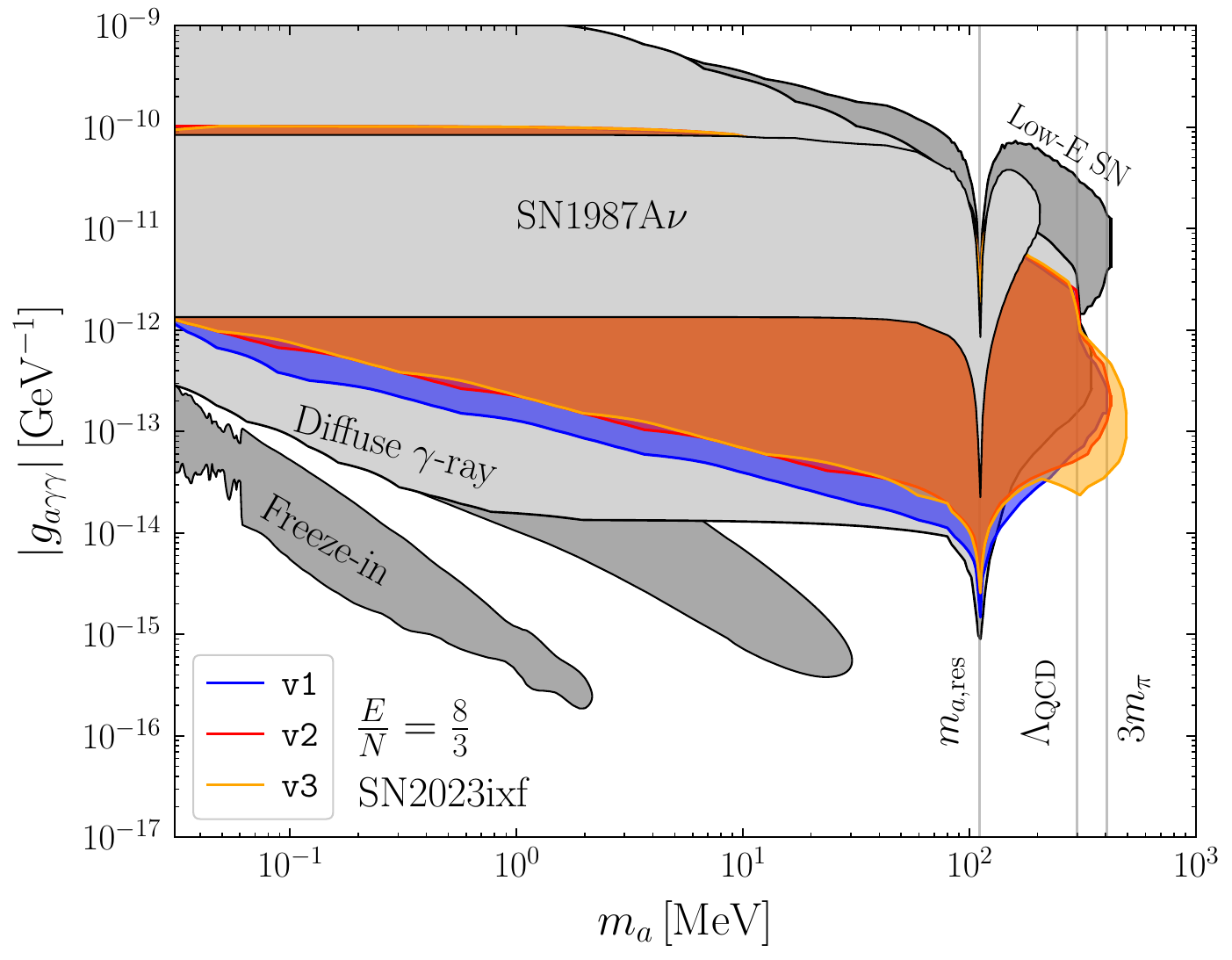}
    \includegraphics[width=0.48\textwidth]{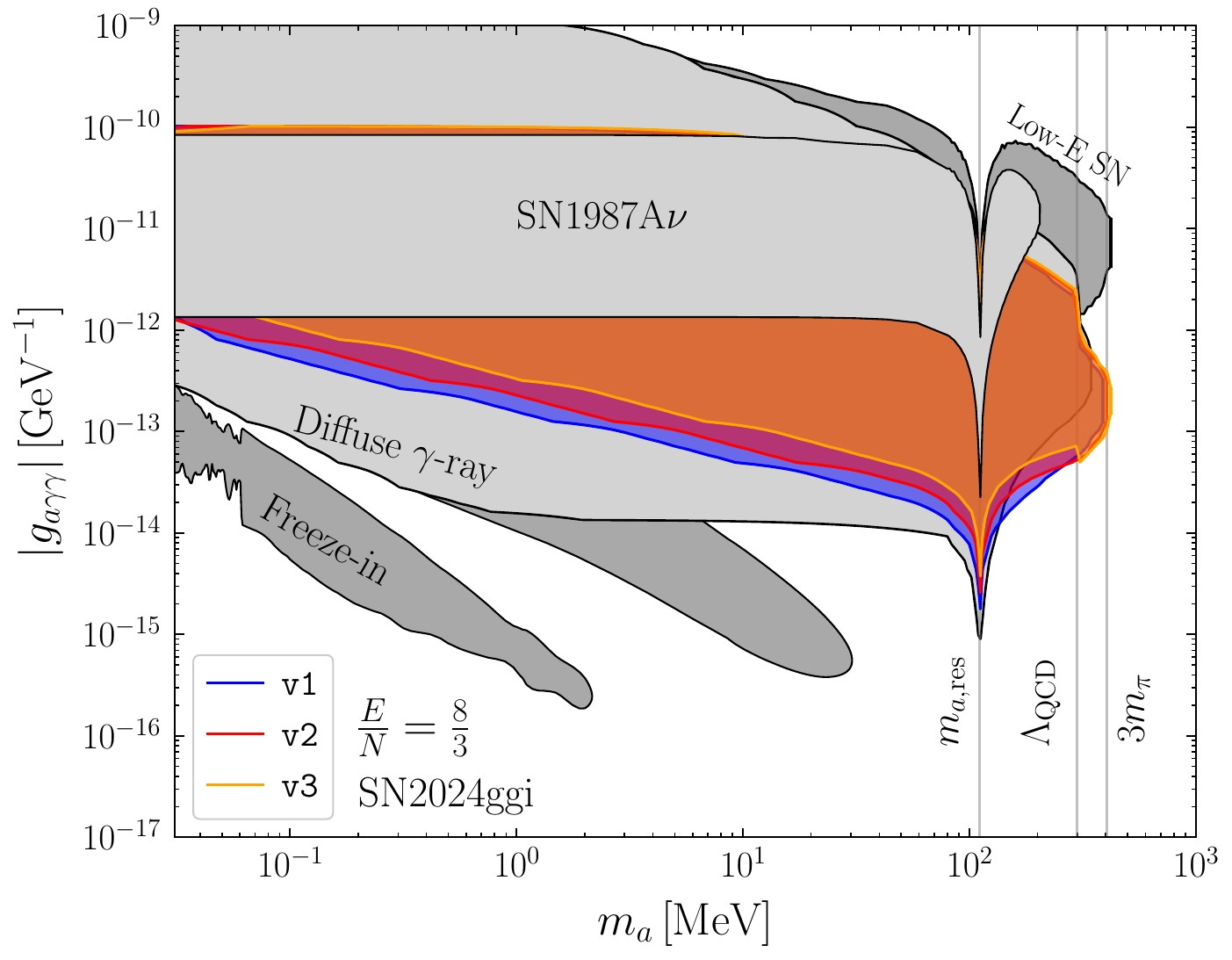}
    \caption{ 
    (\textit{Left}) For SN2023ixf, comparison of limits obtained from the $\texttt{v1}$, $\texttt{v2}$, $\texttt{v3}$  data sets individually. Recall that our fiducial limit excludes any region that is excluded by our analyses of these individual data sets.  (\textit{Right}) As in the left panel but for SN2024ggi.
    }
    \label{fig:limits_v1_v2_v3_2023ixf}
\end{figure}

\begin{figure}[!htb]
    \centering 
    \includegraphics[width=0.5\linewidth]{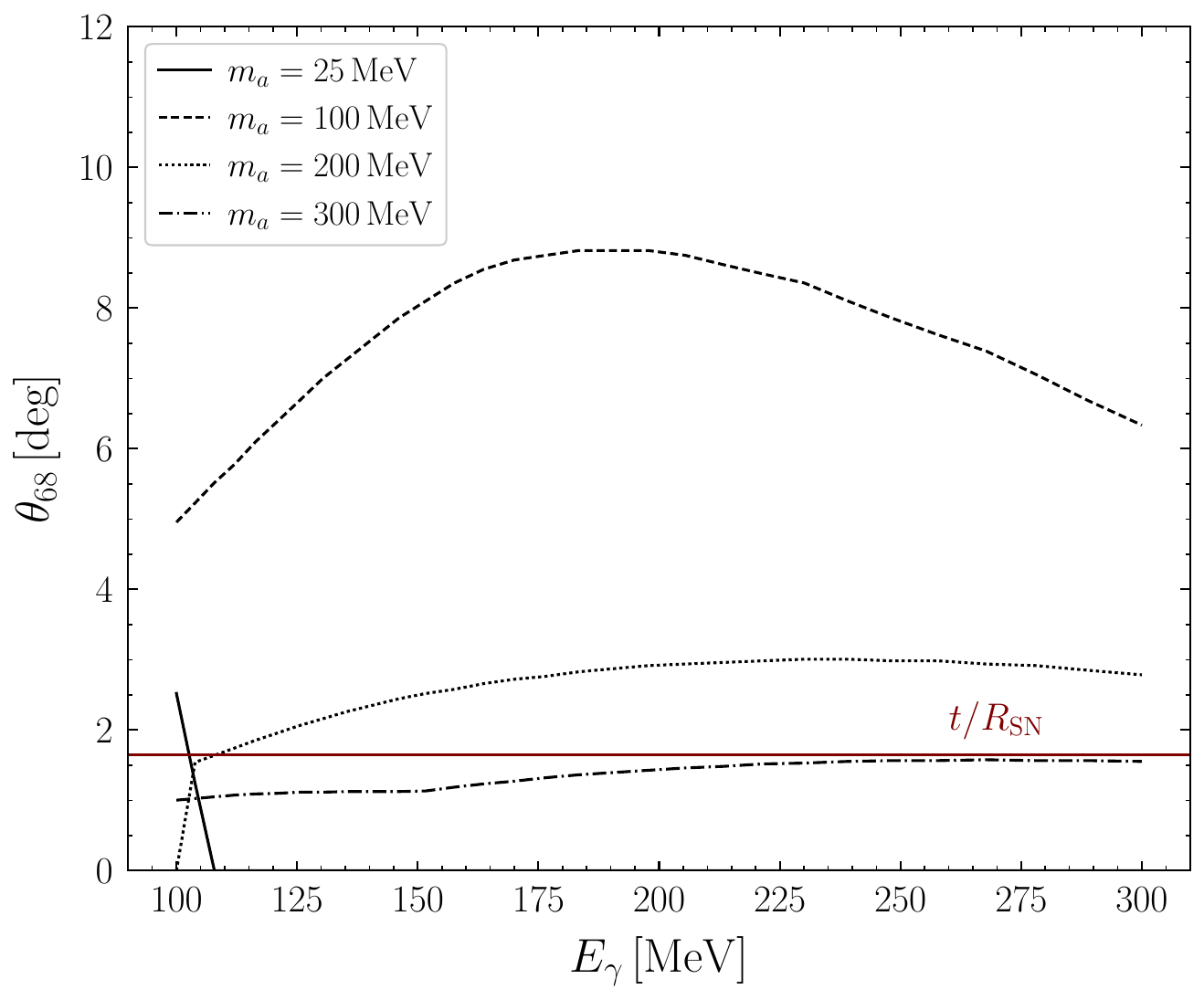}
    \caption{The $68\%$ containment radius $\theta_{68}$ of the decaying axion flux from Cas A, without accounting for the PSF.  This angular spread arises from the fact that the axions may propagate in directions away from the Earth and then decay into photons on Earth-bound trajectories. Given the long baseline from the SN Cas A until today the axion emission from this SN is noticeably extended on the sky for certain axion masses.   
    Here we assume the SFHo-20.0 SN model. Note that the angular spread of the signal is  $\mathcal{O}( t/R_\mathrm{SN})$, where $t\sim 330 $ years is the time delay.
    }   
    \label{fig:Cas_A_angular_spread}
\end{figure}

\begin{figure*}[!t]
    \centering
    \includegraphics[width=0.48\textwidth]{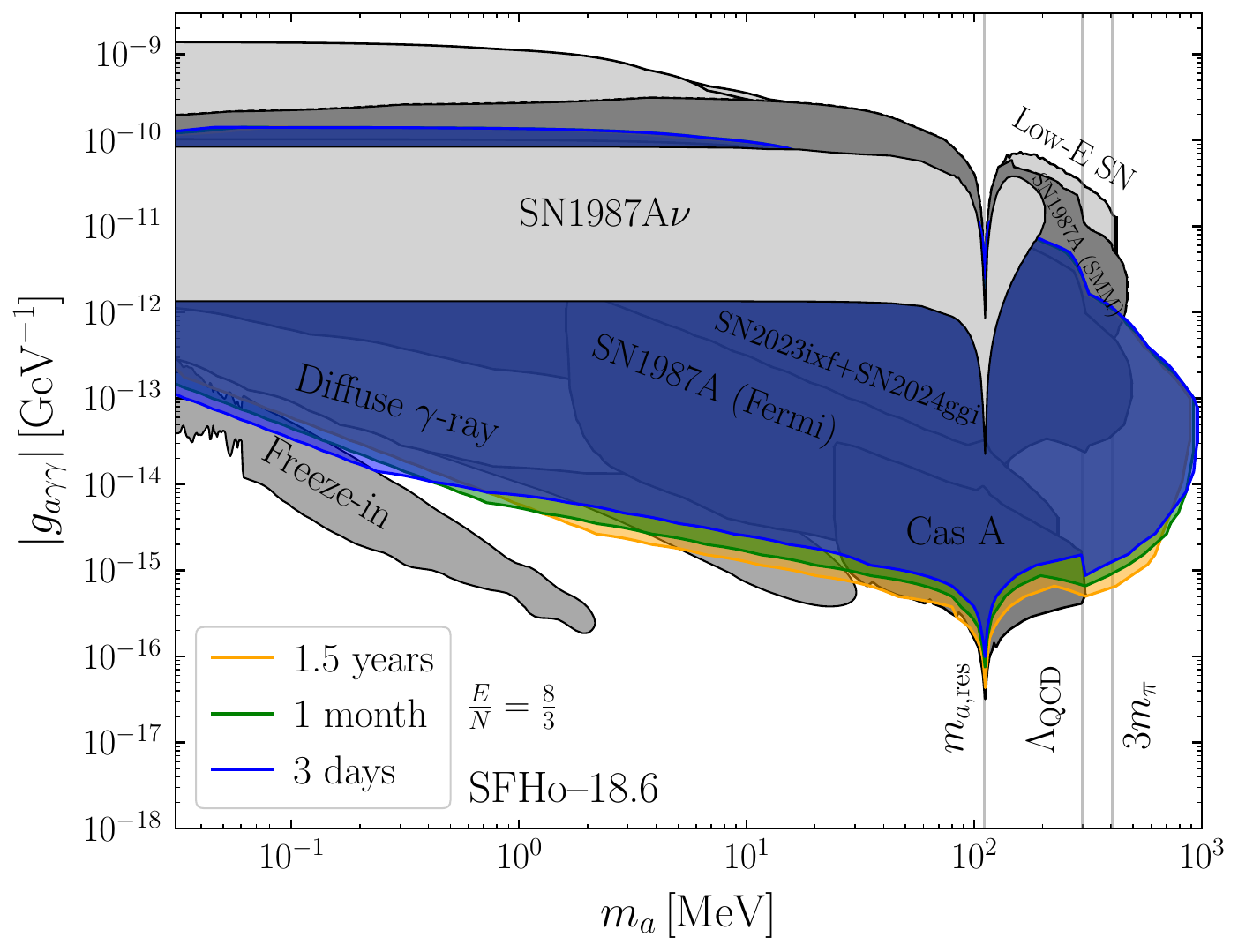 }     \includegraphics[width=0.48\textwidth]{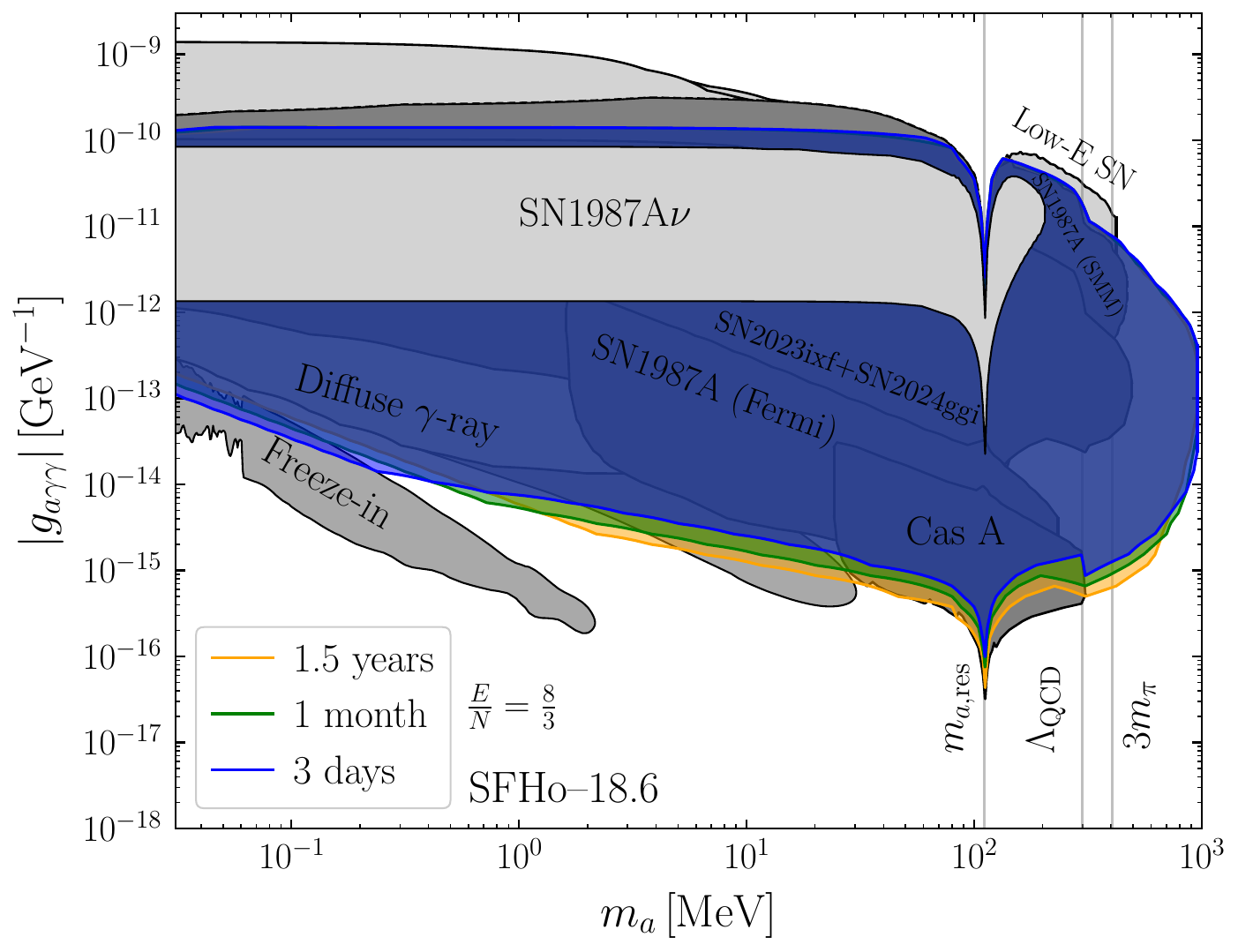}
        \caption{ 
        (\textit{Left}) Projected exclusion regions (at 95\% confidence or greater)  assuming the GUT benchmark model for a future Galactic SN at $10$ kpc from Earth. We generate the null, Asimov Fermi-LAT data ({\it e.g.},~\cite{Safdi:2022xkm}) under the null hypothesis using the best-fit null-hypothesis fit from the relevant analyses of SN2023ixf. We assume the radius of the photosphere is the same as for SN1987A, {\it i.e.} $R_\mathrm{ps}=10^{12}$ m. We project constraints varying the window over which  `ON' data is taken  between 3 days, 30 days, and 1.5 years (as for SN2023ixf). Precisely, we assume that same exposure and analysis framework as in the SN2023ixf \texttt{v1}. \texttt{v2}, and \texttt{v3} data sets. We assume that Fermi is unable to observe the SN during one full orbital period ($\sim 90$ minutes) after the onset of the SN. 
        (\textit{Right}) As in the left panel, but assuming there is no gap between the onset of the SN and the beginning of the observation. Here we assume the SFHo-18.6 SN simulation and shade the existing constraints from this work. Note by construction we do not show any sensitivity for $m_a \ge 1\, \mathrm{GeV}$ as we impose that the axion emission spectrum has no support at energies $E_a \ge 1\, \mathrm{GeV}$, where the chiral perturbation theory assumed in this work is no longer valid.
    }
    \label{fig:projections}
\end{figure*}

\begin{figure}
    \centering
    \includegraphics[width=0.48\textwidth]{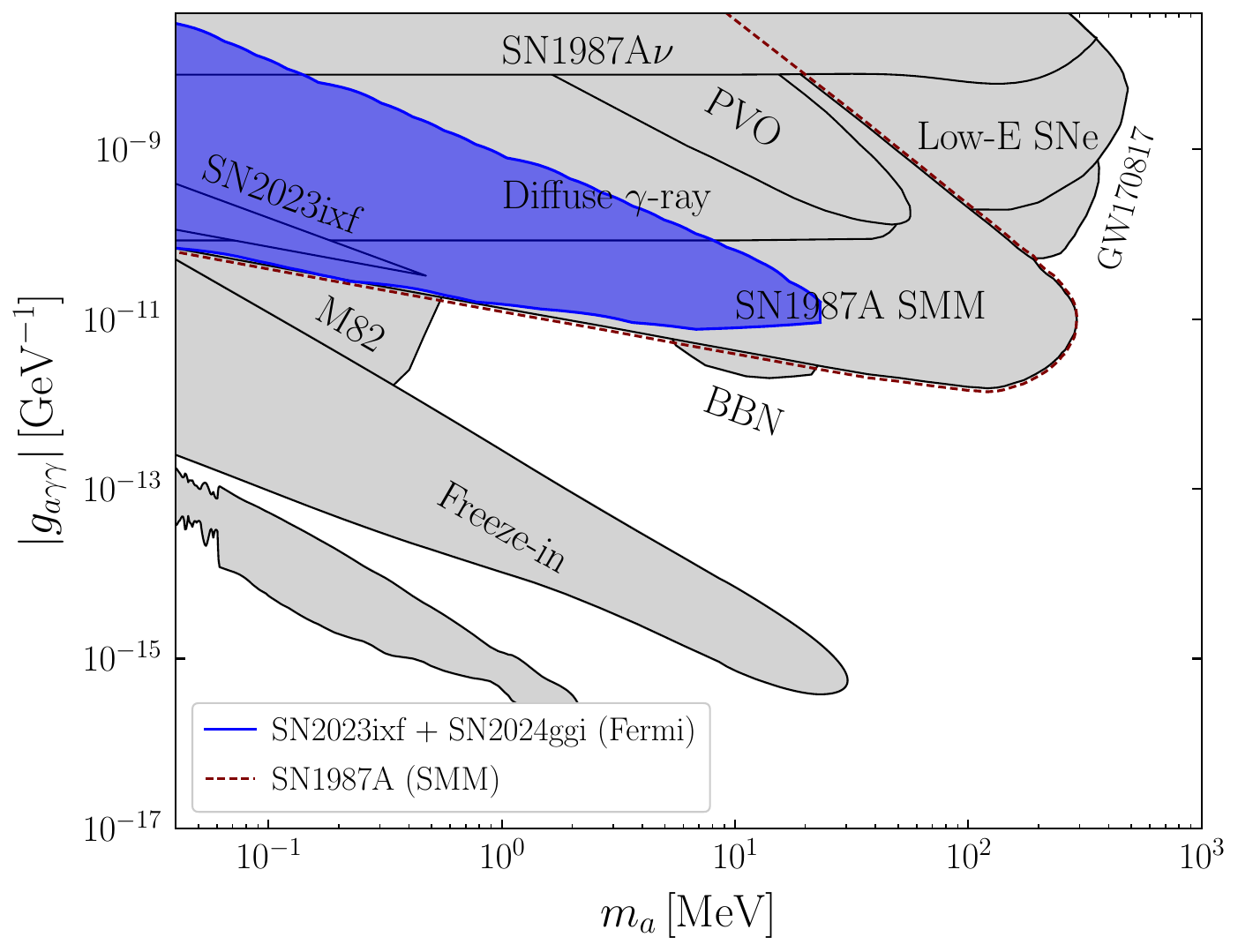}
 \caption{ 
    Summary of constraints derived in this work, as in Fig. \ref{fig:gayy_constraints}, but assuming axions couple to photons alone, though we emphasize that this is not a natural model expectation for heavy axions for the reasons emphasized in the text.  Constraints established from SMM data for decaying axions produced in SN1987A, assuming axion production occurs only through photon conversion and Primakoff processes, are shown in gray shaded~\cite{Jaeckel:2017tud,Hoof:2022xbe,Muller:2023vjm,Lella:2024dmx}.  We reproduce these results in dashed maroon. We account for loop-induced nuclear production processes, though these do not significantly affect the result. In this case we are unable to probe long enough axion lifetimes for the Cas A and SN1987A data to be informative. Instead, only the SN2023ixf and SN2024ggi data lead to new constraints. We note that our constraints improve over previous constraints using SN2023ixf Fermi-LAT data~\cite{Muller:2023pip} (as shown) because we use more relaxed data selection criteria, which gives a smaller time gap between the SN explosion and the onset of data collection (see text for details).   
    }
    \label{fig:limits_summary_plot_photon_couplings}
\end{figure}

We include the following supplementary figures and tables:
\begin{itemize}
    \item Fig.~\ref{fig:gapp_constraints}: Summary of our constraints as a function of $m_a$ versus the axion-proton coupling and versus $1/f_a$.
    \item Fig.~\ref{fig:SN_spectra}: Example axion spectra from SN with and without axion nucleon couplings. 
    \item Fig.~\ref{fig:fermi_PSF}: The Fermi effective area and PSF.
    \item Fig.~\ref{fig:2023_2024_v123}: The spectra extracted from this work for SN2023ixf and SN2024ggi for the individual \texttt{v1}, \texttt{v2}, and \texttt{v3} data sets. 
    \item Fig.~\ref{fig:residual_data}: Example spatial maps of the background-subtracted data for SN2023ixf and SN2024ggi.
    \item Fig.~\ref{fig:data_1987A}: Spatial maps of the data and model components for SN1987A.
    \item Fig.~\ref{fig:Cas_A}: Spatial maps of the data and model components for Cas A.
    \item Fig.~\ref{fig:varying_SN_sim_1987A}: Sensitivity of our results to the SN simulations.
    \item Fig.~\ref{fig:comparison_pion_conversion}: Sensitivity of our SN1987A Fermi result to the inclusion of the axion-pion conversion processes.
    \item Fig.~\ref{fig:limits_v1_v2_v3_2023ixf}: SN2023ixf and SN2024ggi results for the individual \texttt{v1}, \texttt{v2}, and \texttt{v3} data sets.
    \item Fig.~\ref{fig:Cas_A_angular_spread}: Expected angular spread of the signal for Cas A as a result of the long time duration since the SN.
    \item Fig.~\ref{fig:projections}: Projected sensitivity from Fermi-LAT observations of the next Galactic SN.
    \item Fig.~\ref{fig:limits_summary_plot_photon_couplings}: Our results for the axion model where the axion only couples to QED and not to QCD.
    \item Tab.~\ref{tab:properties}: Properties of the SN considered in this work.
    \item Tab.~\ref{tab:time_ranges_Fermi}: The time ranges of Fermi-LAT data incorporated into this work.
    \item Tab.~\ref{tab:best_fits_summary}: A summary of the discovery TSs and best-fit values from the analyses of Fermi-LAT data in this work.
\end{itemize}

\begin{table*}[!htb]{
\centering 
\begin{tabular}{l|l|l|l|l|l|l|l}
\text{Supernova} & \text{Optical Signal (UTC)} & $R_\mathrm{SN}$ [$\mathrm{kpc}$] &  $R_\mathrm{ps}$ [$R_{\odot}$] & $M_\mathrm{proj} \, [M_\odot]$ &  $M_\mathrm{NS} \, [M_\odot]$ & Refs. &   \\
\hline 
\hline 
SN1987A & 1987-02-24 (Las Campanas Observatory) &  $51.4 \pm 1.2$ & $\approx 43$ & $\sim 18-20$  & $1.22 - 1.62$ & \cite{Kazanas:2014mca,Panagia:2003rt, 1988ApJ...330..218W, Page:2020gsx,Utrobin:2018mjr}  \\
Cas A & $\approx 1674$  &  $ 3.4_{-0.1}^{+0.3} \,\mathrm{kpc}$ & $\quad\,\,\,$--- & $\sim 15-25$  & $1.65 - 2.01$  & \cite{Milisavljevic:2024mbg, Heinke_2010, Bonanno:2013oua}  \\ %
SN2023ixf & 2023-05-18 19:30:00.000 (Kōichi Itagaki)  & $(6.85 \pm 0.15)\times 10^{3}$&  $410 \pm 10$ & $11 \pm 2$ & $\quad\quad$--- & \cite{Hosseinzadeh:2023ixa,2023ATel16050....1S, Pledger:2023ick,Kilpatrick:2023pse, 2023TNSAN.130....1M}  \\
SN2024ggi & 2024-04-11 03:22:35.616 (ATLAS) & $(6.72 \pm 0.19)\times 10^{3}$& $887 ^{+60}_{-51}$  & $13 \pm 1$ & $\quad\quad$--- & \cite{Chen:2024mye, Chen:2024mkm, Xiang:2024nee, tns_2024ggi} \\
\end{tabular}}
\caption{
\label{tab:properties} Properties of the SN studied in this work.  For each SN we indicate the time (in UTC if known) of the optical signal at Earth and the observatory (or observer, in the case of SN2023ixf) which first received it, the distance $R_\mathrm{SN}$ to Earth, the radius of the photosphere $R_\mathrm{ps}$, the progenitor mass $M_\mathrm{proj}$, and the mass of the resulting neutron star $M_\mathrm{NS}$. Note that for some quantities estimates do not exist in the literature, in which case we leave a blank.}
\end{table*}

\begin{table*}[!htb]
\begin{tabular}{l|c|c}
\text{Data set} & \text{Start Time (s)} & End Time (s)   \\
\hline 
\hline 
SN1987A & 239557417  &  741142542 \\
\hline 
Cas A & 239557417  &  741142542 \\
\hline
SN2023ixf (\texttt{bkg}) & 239557417 & 705958205 \\
SN2023ixf (\texttt{v1}) & 706044605 & 754856440 
\\
SN2023ixf (\texttt{v2}) & 706044605 & 708723005 \\
SN2023ixf (\texttt{v3}) & 706044605 & 706390205 \\
\hline
SN2024ggi (\texttt{bkg}) & 239557417 & 734325760 \\
SN2024ggi (\texttt{v1}) & 734412160 & 754856440 
\\
SN2024ggi (\texttt{v2}) & 734412160 & 737090560 \\
SN2024ggi (\texttt{v3}) & 734412160 & 734757760 \\
\end{tabular}
\caption{\label{tab:time_ranges_Fermi} Time ranges of Fermi-LAT data used in this work in mission elapsed time (MET).
}
\end{table*}

\begin{table*}[!htb]
\centering 
\begin{tabular}{l|c|c}
\text{Analysis} & Best-fit $(m_a \,[\mathrm{MeV}] \, , |g_{a\gamma\gamma}|\, [\mathrm{GeV}^{-1}])$ & Discovery TS   \\
\hline 
\hline 
SN1987A &  (234  , $5.3 \times 10^{-15}$)  & 6.4  \\
\hline
Cas A (point source) &  (50  , $-2.3\times 10^{-16}$) & 0  \\
\hline
SN2023ixf (\texttt{v1}) & (405  , $-3.2 \times 10^{-13}$) & 0 \\
SN2023ixf (\texttt{v2}) & (164 , $2.0\times 10^{-14}$) & 3.0 \\
SN2023ixf (\texttt{v3}) & (17  , $7.1 \times 10^{-11}$) &  2.4\\
\hline
SN2024ggi (\texttt{v1}) & ($1.9 \times 10^{-2}$ , $1.0\times 10^{-10}$) & 4.74 \\
SN2024ggi (\texttt{v2}) & (6.7 , $1.0 \times 10^{-10}$) & 0.86
 \\
SN2024ggi (\texttt{v3}) & (422  , $4.6 \times 10^{-13}$) & 0.44 \\
\hline
SN2023ixf + SN2024ggi (\texttt{v1})  & ($1.4 \times 10^{-2}$, $1.5 \times 10^{-10}$) & 1.0 \\
SN2023ixf + SN2024ggi (\texttt{v2}) & (171 , $2.0 \times 10^{-14}$) & 2.3 \\
SN2023ixf + SN2024ggi (\texttt{v3})  & (12, $8.5\times 10^{-11}$) & 2.8 \\
\end{tabular}
\caption{Summary of best-fit points and associated discovery test statistics (TS) for each of the analyses of Fermi-LAT data conducted in this work. Note that for Cas A we indicate these quantities for the analysis assuming no spatial extension of the signal
beyond the PSF.
\label{tab:best_fits_summary}}
\end{table*}

\FloatBarrier

\newpage
\section{Additional decay channels}
\label{sec:electron_decays}
For $m_a > 3 m_\pi$, the axion may decay into three pions with the rate \cite{Bauer:2017ris}
\begin{align}
\Gamma_{a 3 \pi}= \frac{m_a m_\pi^4\left(\Delta C_{u d}\right)^2}{6144 \pi^3 f_\pi^2 f_a^2} &\Theta\left(m_a-3 m_\pi\right)\left[g_0\left(\frac{m_\pi^2}{m_a^2}\right)+g_1\left(\frac{m_\pi^2}{m_a^2}\right)\right] \,,
\label{eq:a2yy_rate}
\end{align}
where $\Delta C_{u d}=C_{auu}-C_{add}+ \frac{m_d-m_u}{m_d+m_u}$ and
\begin{align}
g_n(r)= \frac{2 \cdot 6^n}{(1-r)^2}& \int_{4 r}^{(1-\sqrt{r})^2} d z \sqrt{1-\frac{4 r}{z}}(z-r)^{2 n} \sqrt{1+z^2+r^2-2 z-2 r-2 z r} \,.
\end{align}
In our computation of the signal flux, we conservatively assume that decays into pions do not produce visible photons in the detector, although in reality they would. As we show in Fig. \ref{fig:axion_lifetime}, compared to the case where only decays into photons are accounted for, including pion decays as described decreases the axion lifetime  by an $\mathcal{O}(1)$ factor for $m_a > 3 m_\pi$. As the signal flux is exponentially sensitive to the lifetime, this has a non-negligible effect on our $g_{a\gamma\gamma}$ constraints. 

For $m_a > 2m_e$, depending on the details of the axion UV completion, we may also have decays into electrons through the interaction
\begin{equation}
\mathcal{L}=C_{a e e} \frac{\partial_\mu a}{2 f_a} \bar{e} \gamma^\mu \gamma_5 e \,,
\end{equation}
with decay rate \cite{Foster:2022ajl}
\begin{align}
\Gamma_{a \rightarrow e^{+} e^{-}} & =\frac{m_a m_e^2}{8 \pi f_a^2} C_{a e e}^2\left[1-4 \frac{m_e^2}{m_a^2}\right]^{1 / 2} \,, 
\end{align}
such that for $m_a \gg m_e$,
\begin{align}
\frac{\Gamma_{a\to ee} }{\Gamma_{a \to \gamma\gamma}}  \approx 5.9 \times 10^6 \left(\frac{0.511 \mathrm{MeV}}{m_a}\right)^2\left(\frac{C_{aee}}{C_{a\gamma\gamma}}\right)^2
\,.
\label{eq:electon_decay_Br}
\end{align}
In the minimal GUT scenario with no axion-electron coupling at tree-level, the latter is generated at 1-loop. In this case we find $C_{aee}/C_{a\gamma\gamma} \sim 5\times 10^{-5}$, such that the electron decay channel is suppressed relative to decays into photons.

\section{Heavy Axions and the strong-{\it CP} problem}
\label{sec:heavy_axion}

In this work we focus on heavy axions that couple to both QED and QCD but acquire their dominant mass contributions through non-Standard Model sectors. In particular,   we consider axion-like particles which in the UV couple to Standard Model gauge bosons,
\begin{equation}
\label{eq:Lag_1}
    \mathcal{L} \supset -\frac{1}{4}\frac{E_{W}}{N} \frac{g_2^2}{8\pi^2 f_a} a\, W_{\mu\nu}^a\tilde{W}^{a\, \mu\nu} - \frac{1}{4}\frac{E_{B}}{{N}} \frac{g_1^2}{8\pi^2 f_a} a\, B_{\mu\nu}\tilde{B}^{\mu\nu} - \frac{1}{4} \frac{g_s^2}{8\pi^2 f_a} a\, G_{\mu\nu}\tilde{G}^{\mu\nu}\,,
\end{equation}
with $W_{\mu \nu}$, $B_{\mu \nu}$ and $G_{\mu \nu}$ the $\mathrm{SU}(2)_L$, $\mathrm{U}(1)_Y$ and $\mathrm{SU}(3)_C$ field strengths, respectively.  
Note that axion-matter couplings are then generated under the RG flow ~\cite{Srednicki:1985xd,Bauer:2017ris}. For ``standard" GUT embeddings of the Standard Model, the coefficients $E_{W}$ and $E_{B}$ are related to the anomaly coefficients of QED and QCD by $E_{W} =N$ and $E_{B} = E - N$.  In addition to the Lagrangian above, we assume that the axion acquires a mass through instantons of a hidden sector or through {\it e.g.} string theory or gravitational instantons, as happens in the axiverse.  Let us denote the dominant mass contribution to the axion potential by 
\es{eq:high_scale}{
{\mathcal L} \supset \Lambda^4 \left(1 - \cos\left({a \over f_a } - \theta_1\right) \right) \,,
}
where the precise form of the potential is not important so long as it periodic with period $2 \pi f_a$.  The axion mass is then $m_a = \Lambda^2 / f_a$.  

Below the QCD confinement scale non-perturbative QFT effects generate a contribution to the axion's potential of the form 
\es{eq:QCD_pot}{
{\mathcal L} \supset \chi_{\rm top} \left(1 -  \cos\left({ N a \over f_a} - \bar \theta_{\rm QCD}\right) \right) \,,
}
where $\chi_{\rm top} \approx (75.5 \, \, {\rm MeV})^4$ is the topological susceptibility~\cite{diCortona:2015ldu} and where we have approximated the form of the potential by a simple cosine since the details of the full potential are not important for the argument in this section.  We focus on heavy axions in this work with masses in the MeV - GeV range, implying that we assume $\Lambda^4 \gg \chi_{\rm top}$. This, in turn, implies that when the axion minimizes its potential it induces an effective $\theta$-angle for QCD, which would contribute to the neutron electric dipole moment: 
\es{}{
\langle {N a \over f_a} - {\bar \theta_{\rm QCD}} \rangle \approx N \theta_1 - \bar \theta_{\rm QCD} \,.
}
Generically, the two angles are not aligned and the heavy axion does not solve the strong-{\it CP} problem (but see Refs.~\cite{Rubakov:1997vp,Berezhiani:2000gh}). 

While the heavy axion does not solve the strong-{\it CP} problem, we note that the heavy axion allows for the existence of a normal QCD axion which does.  To see this, let us now add a second axion to the theory. In particular, we allow for a second axion $\tilde a$ that couples to the SM through the same operators as in~\eqref{eq:Lag_1} but with a color anomaly coefficient $\tilde N$. (The axion $\tilde a$ could also have a different $\tilde E$, but this will not be relevant for the discussion.) We further assume that both axions receive potential contributions from the hidden-sector instantons, such that -- without loss of generality -- we may modify~\eqref{eq:high_scale} to
\es{eq:high_scale_2}{
{\mathcal L} \supset \Lambda^4 \left(1 - \cos\left({a \over f_a } + {\tilde M} {\tilde a \over f_a}- \theta_1\right) \right) \,,
}
where ${\tilde M}$ is a positive integer.  On the other hand, the QCD-induced potential now becomes
\es{eq:QCD_pot_2}{
{\mathcal L} \supset \chi_{\rm top} \left(1 -  \cos\left({ N a \over f_a} + {\tilde N a \over f_a} - \bar \theta_{\rm QCD}\right) \right) \,.
}
Minimizing the full potential, we see that there are two angle $\theta_1$ and ${\bar \theta}_{\rm QCD}$ but also two dynamical degrees of freedom, $a$ and $\tilde a$, such that at the minimum of the potential:
\es{}{
\langle \chi_{\rm top} { N a \over f_a} + {\tilde N a \over f_a} - \bar \theta_{\rm QCD} \rangle = 0 \,.
}
Thus, the strong-{\it CP} problem is precisely solved in this case. We may go one step further and diagonalize the mass matrix to identify a heavy axion state and a lighter, normal QCD axion state with mass $m_{a, {\rm QCD}} \approx \chi_{\rm top}^{1/2} / f_a$.  The heavy state is 
\es{}{
a_{\rm heavy} \propto a + \tilde M \tilde a \,,
}
while the lighter, QCD axion state is 
\es{}{
a_{\rm QCD} \propto -\tilde M a + \tilde a \,,
}
to leading order in $\chi_{\rm top} / \Lambda^4$.  The phenomenology of the QCD axion state is the same as usual for a QCD axion.  The heavy axion, $a_{\rm heavy}$, couples to both QCD and to QED.

In the presence of a third instanton potential, there is an additional phase that cannot be removed by the two axion states. (More generally, if there are $N$ axions and $M$ instanton contributions, with $M > N$, then there are $M - N$ physical phases.)  However, this is no different from the normal story of a QCD axion with an additional contribution to its potential; we may disrupt the solution to the strong-{\it CP} problem unless the scale of the additional potential is sufficiently far below that of $\chi_{\rm top}^{1/4}$. In particular, if there is an additional instanton contribution with scale $\Lambda_3$, we -- generically -- need $\Lambda_3 / \chi_{\rm top}^{1/4} \lesssim 0.05$ to not spoil the QCD axion solution to the strong-{\it CP} problem. 

It is important to contrast the heavy axion discussion above with that of ultra-light axions that also couple to QCD. While heavy axions ($m_a \gg \chi_{\rm top}^{1/2}/f_a$) with non-trivial couplings to QCD are generic and expected, light QCD axions (with mass $m_a \ll \chi_{\rm top}^{1/2}/ f_a$) that couple to QCD are strongly fine tuned or require additional model-building~\cite{Hook:2018jle,Gomez-Banon:2024oux}. 

\section{Axion-matter couplings under RG flow}
\label{sec:RGflow}

After integrating out the heavy fields at the electroweak symmetry breaking scale in~\eqref{eq:Lag_1}, we obtain the effective Lagrangian
\begin{equation}
    \mathcal{L} \supset - \frac{1}{4}\frac{E}{N} \frac{e^2}{8\pi^2 f_a} a\, F_{\mu\nu}\tilde{F}^{\mu\nu} - \frac{1}{4} \frac{g_s^2}{8\pi^2 f_a} a\, G_{\mu\nu}\tilde{G}^{\mu\nu}\,,
\end{equation}
where $F_{\mu\nu}$ is the electromagnetic field strength, and matching imposes $E = E_{W} + E_{B}$. At scales below the QCD confinement scale, $\Lambda_{\rm QCD}$, \mbox{$g_{a\gamma\gamma}$} receives additional contributions due to the axion-pion mixing, such that we define $g_{a\gamma\gamma} \equiv \frac{\alpha_{\rm EM}}{2 \pi f_a}\,C_{a\gamma\gamma}$, with ~\cite{Bauer:2021mvw}
\begin{align}
\begin{split}
C_{a\gamma\gamma}(\mu > \Lambda_{\rm QCD}) &= \frac{E}{N}\\
C_{a\gamma\gamma}(\mu < \Lambda_{\rm QCD}) &\simeq \frac{E}{N} -1.92\, -\frac{m_a^2}{m_\pi^2-m_a^2}\left[ \frac{m_d-m_u}{m_d+m_u}\, +\frac{C_{au u}(2\, {\rm GeV})-C_{ad d}(2\, {\rm GeV})}{2}\right] \,,
\end{split}
\label{eq:GUT_relation_cagg}
\end{align}
where the axion-quark couplings are loop-induced in the scenarios considered in this work and this result is valid for $\left|m_\pi^2-m_{a}^2\right| \gg m_\pi^2 f_\pi / f_a$. Thus, the regime where \eqref{eq:GUT_relation_cagg} becomes invalid is an interval around $m_\pi$ whose width is too small to be visible on our constraint plots.  

\begin{figure}
    \centering
    \includegraphics[width=0.5\linewidth]{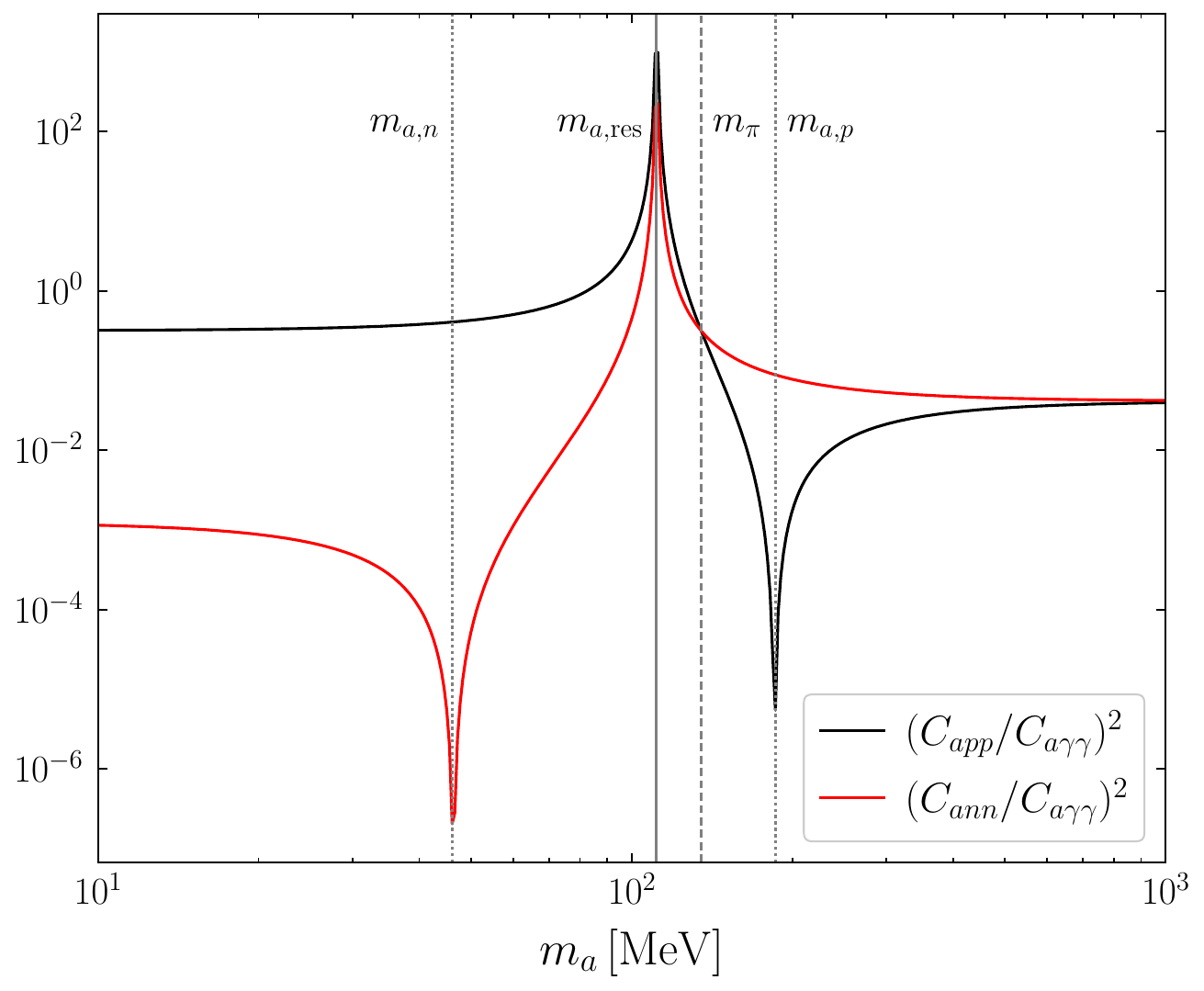}
    \caption{Nucleon couplings in the GUT model ($E/N = 8/3$ with no tree-level axion-quark couplings in the UV) for $g_{a\gamma\gamma}=10^{-13} \, \mathrm{GeV}^{-1}$. Note that here we consider the IR value of $g_{a\gamma\gamma}$, {\it i.e.} at energy scales below $\Lambda_\mathrm{QCD}$. At large $m_a$ we have $C_{app}=C_{ann}$, and in the limit $m_a \to m_\pi$ we have $C_{app}=-C_{ann}$. As discussed in the text, the axion-photon coupling vanishes at $m_a=m_{a,\mathrm{res}}$. Further, for each nucleon there is a unique value of $m_a$ for which its axion coupling vanishes (denoted above as $m_{a,n}$ ($m_{a,p}$) for neutrons (protons)). 
}
    \label{fig:nucleon_couplings_GUT}
\end{figure}

Note that for a given choice of $E/N$ there is a value of $m_{a}$ for which $C_{a\gamma\gamma}$ vanishes:
\es{}{
m_{a,\mathrm{res}}  \approx m_\pi\left[1 + \frac{m_d-m_u}{m_d+m_u} \left(E/N - 1.92\right)^{-1} \right]^{-\frac{1}{2}}\,.
}
For our benchmark scenario of $E/N = 8/3$, $m_{a,\mathrm{res}} \approx 0.82 m_\pi$. Above, we neglect the difference $C_{auu}-C_{add}$ in the bracketed term of \eqref{eq:GUT_relation_cagg} as it is subdominant.
Referring to {\it e.g.} Fig.~\ref{fig:gayy_constraints}, we are able to probe small values of $|g_{a\gamma\gamma}|$ for $m_a \approx m_{a, {\rm res}}$. We may understand this behavior through the following simple scaling relations. For $m_a \approx m_{a, {\rm res}}$, the axion-photon coupling scales with mass as $g_{a\gamma\gamma} \propto (m_a^2 - m_{a,{\rm res}}^2) / f_a$ while the axion-nucleon couplings are $m_a$-independent to first approximation; {\it e.g.}, $g_{app} \propto 1/f_a$. Since the observed flux for long axion lifetimes scales as $g_{a\gamma\gamma}^2 g_{app}^2$, where $g_{app}$ is a proxy for the ensemble of axion nucleon couplings, this implies that for a flux upper limit $F_{\rm U.L.}$ the 95\% lower limit on $f_a$, which we refer to as $f_a^{95}$, scales with $m_a$ as $f_a^{95} \propto (m_a^2 - m_{a,{\rm res}}^2)^{1/2} / F_{\rm U.L.}^{1/4}$. Thus, as $m_a \to m_{a,{\rm res}}$, the 95\% lower limit on $f_a$ goes to zero, meaning that we lose the ability to constrain $f_a$. This behavior is clearly seen in Fig.~\ref{fig:gapp_constraints}.  On the other hand, in the vicinity of $m_{a,{\rm res}}$ this implies that the 95\% upper limit on $g_{a\gamma\gamma}$, which we refer to as $g_{a\gamma\gamma}^{95}$, scales with $m_a$ as $g_{a\gamma\gamma}^{95} \propto F_{\rm U.L.}^{1/4} \sqrt{m_a^2 - m_{a,{\rm res}}^2}$, implying that near $m_{a,{\rm res}}$ the upper limit $g_{a\gamma\gamma}^{95}$ goes to zero, as seen in Fig.~\ref{fig:gayy_constraints}.

\begin{figure}
    \centering
    \includegraphics[width=0.5\linewidth]{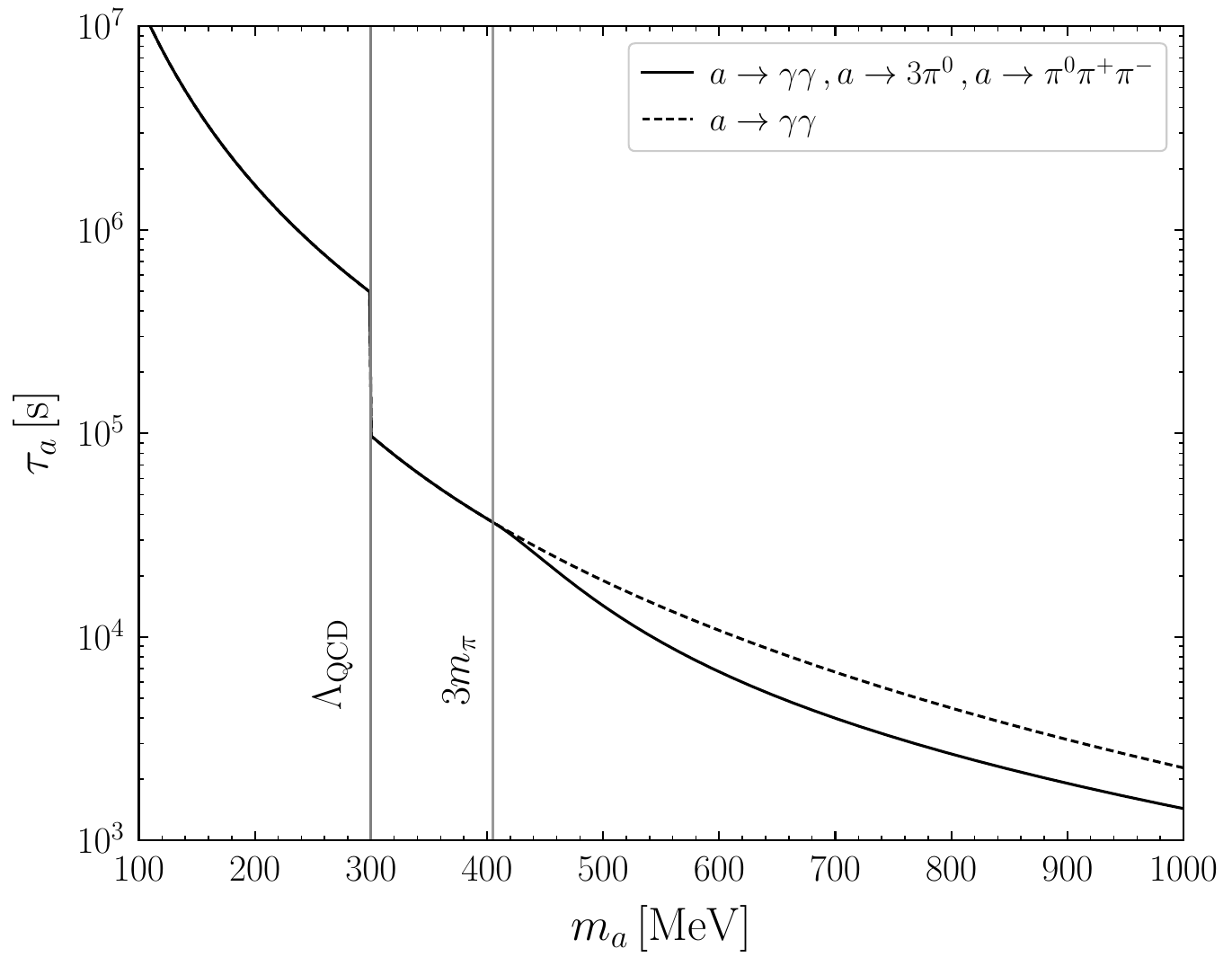}
    \caption{The axion lifetime as a function of the axion mass in the GUT benchmark model, fixing the IR value of the axion-photon coupling to $g^{\mathrm{IR}}_{a\gamma\gamma} = 10^{-13} \, \mathrm{GeV}^{-1}$. We compare the case where only decays into photons are accounted for (solid) to the case where decays into three pions are also included (dashed). Note that the value of $g_{a\gamma\gamma}$ which enters in the lifetime is evaluated at the energy scale $m_a$ and consequently receives a threshold correction for $m_a > \Lambda_\mathrm{QCD}$ (see SM Sec. \ref{sec:RGflow} for details).
}
    \label{fig:axion_lifetime}
\end{figure}

After the axions exit the PNS they decay, at the energy scale given by the center-of-mass energy $m_a$, such that if $m_a \ge \Lambda_\mathrm{QCD}$ the axion-photon coupling that should be used when computing the lifetime via \eqref{eq:a2yy_rate}  differs from the far-IR value $g^\mathrm{IR}_{a\gamma\gamma}$ by the pion contribution. Similar considerations apply when computing the axion emission spectrum for inverse photon decays, while for axion production through the Primakoff process, the center-of-mass energy is dependent on the axion energy $E_a$.\footnote{As the production processes involving nucleons are dominant over Primakoff emission, we simply approximate that $g_{a\gamma\gamma}$ is equal to it's value at the energy scale $m_a$ in the computation of the Primakoff emission spectrum, independently of the axion energy.} This threshold effect on the axion lifetime is shown in Fig. \ref{fig:axion_lifetime} and translates to a kink in our constraints curves at $m_a =\Lambda_\mathrm{QCD}$, visible in {\it e.g.}, Fig.~\ref{fig:gayy_constraints}.

\subsection{Couplings with Nucleons}

The axion-quark and axion-nucleon couplings are obtained via an analogous derivation to that in Ref. \cite{Manzari:2024jns}, where in the RG evolution of the axion-quark couplings we must now account for the contribution from $\mathrm{SU}(3)_C$ in addition to that from $\mathrm{U}(1)_{\rm EM}$.
In the IR  (for ${\max}(m_a,\, \Lambda_{\rm QCD}) < \mu < m_Z$) we obtain
\begin{equation}
    C_{aqq}(\mu) = C_{aqq}(m_Z) + \frac{3}{64\pi^4}\ln\left(\frac{m_Z}{\mu}\right) Q^2e^2(m_Z)e^2(\mu)C_{a\gamma\gamma} + \frac{1}{16\pi^4}\ln\left(\frac{m_Z}{\mu}\right) g_s^2(m_Z)g_s^2(\mu)\,,
    \label{eq:RunningBelowEW}
\end{equation}
where $Q$ is the electromagnetic charge of the quark, and $m_Z$ is the mass of the Z-boson.
After matching, the axion-nucleon couplings (defined in \eqref{eq:Lagnucleons}) are given by
\begin{align}
    \begin{split}
        C_{app} = \frac{g_0}{2}\bigg[C_{auu}(2\, {\rm GeV}) + C_{add}(2\, {\rm GeV}) - 1\bigg] + \frac{g_A}{2}\frac{m_{\pi}^2}{m_{\pi}^2-m_a^2}\bigg[C_{auu}(2\, {\rm GeV}) - C_{add}(2\, {\rm GeV}) - \frac{m_d - m_u}{m_d+m_u}\, \bigg]\,,\\
        C_{ann} = \frac{g_0}{2}\bigg[C_{auu}(2\, {\rm GeV}) + C_{add}(2\, {\rm GeV}) - 1\bigg] - \frac{g_A}{2}\frac{m_{\pi}^2}{m_{\pi}^2-m_a^2}\bigg[C_{auu}(2\, {\rm GeV}) - C_{add}(2\, {\rm GeV}) - \frac{m_d - m_u}{m_d+m_u}\, \bigg]\,,
    \end{split}
    \label{eq:CapCan_wmpipole}
\end{align}
where $m_{d(u)}$ is the down-(up-) quark mass at 2 GeV and the contribution from the axion-pion mixing is evident. Following Ref.~\cite{GrillidiCortona:2015jxo}, we use $g_0 = 0.521(53)$ and $g_A = 1.2723(23)$.\footnote{Note that in Ref. \cite{Lella:2024dmx}, which derives bounds on the axion-nucleon couplings from SN observations, the $m_a$ dependence in \eqref{eq:CapCan_wmpipole} is not accounted for, although we find that it is crucial to include it.}  
Note that the neutron (proton) coupling vanishes at a unique value of $m_a$, which we denote $m_{a,n}$ ($m_{a,p}$). We compute $m_{a,n} \approx 46.1\, \mathrm{MeV}$ and $m_{a,p} \approx 185.9\, \mathrm{MeV}$.

\section{Axion emission spectrum}
\label{sec:absorption}

We compute the axion emission spectra $d^2 n_a/d E_a /dt$ using a suite of 1D simulations in the Garching archive (see main text and  Ref.~\cite{Manzari:2024jns} for details). Due to weak equilibrium in the PNS, the pion chemical potential satisfies $\mu_{\pi^-} = \mu_n - \mu_p$, and when $\mu_{\pi^-} > m_{\pi^-}$ a Bose-Einstein condensate of pions is favored. For axion production through pion conversion  we make the conservative choice of accounting only for the region of the PNS where $\mu_{\pi^-} < m_{\pi^-}$ and thus where there is no Bose-Einstein condensation. 

\begin{figure}[h]
    \centering 
    \includegraphics[width=0.48\textwidth]{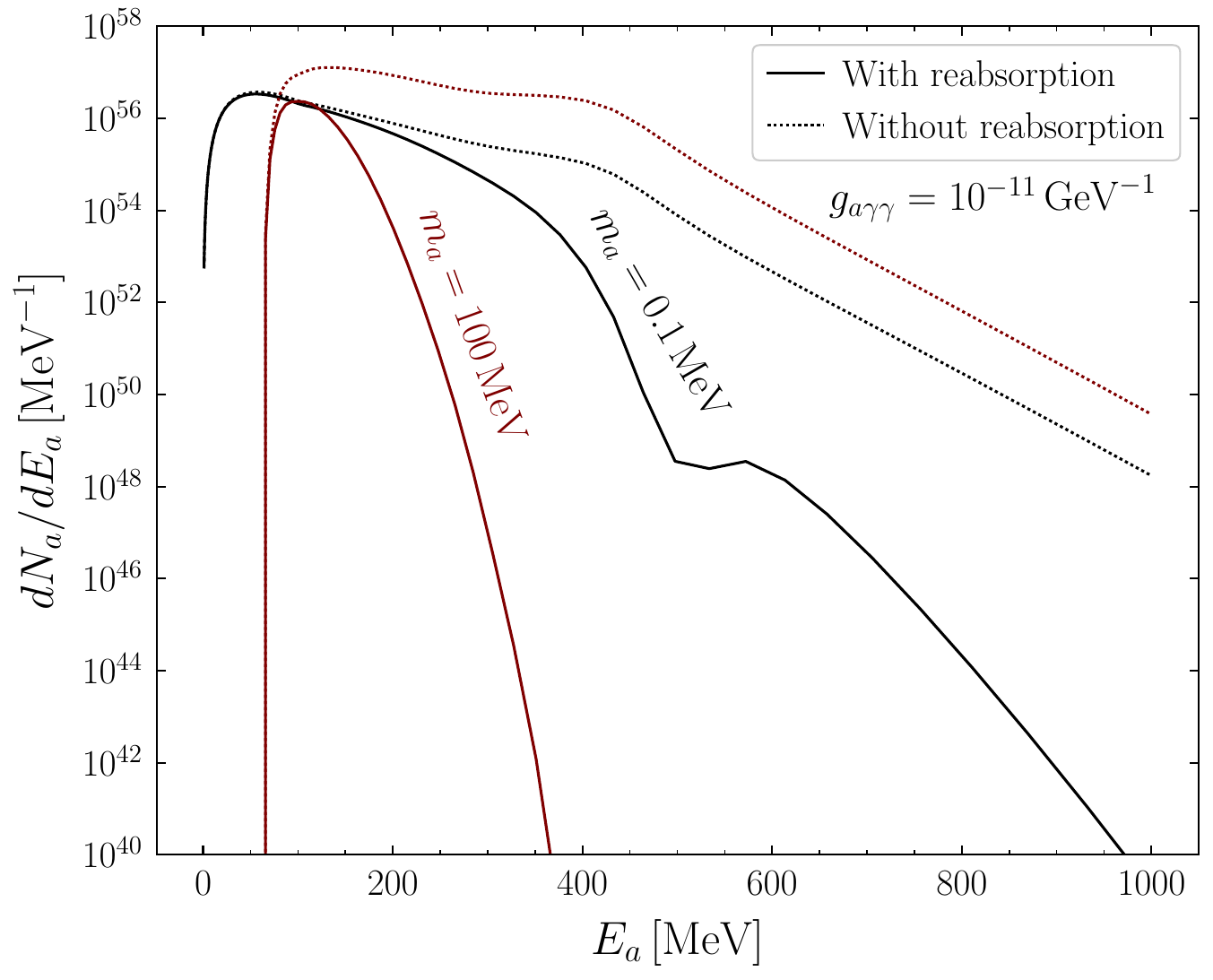}
    \caption{The axion emission spectrum integrated over the first $\sim$$10$ s of the SN, assuming the GUT benchmark model, including  all relevant processes: photon coalescence, Primakoff production, nucleon-nucleon bremsstrahlung, and pion conversion. Here we fix $g_{a\gamma\gamma} = 10^{-11} \, \mathrm{GeV}^{-1}$, and show the spectra for $m_a = 0.1 \, \mathrm{MeV}$ (black) and $m_a = 100 \, \mathrm{MeV}$ (maroon). We compare the spectrum accounting for reabsorption within the PNS (solid) (see text for details), to the result obtained ignoring the reabsorption (dotted). We compute SN properties using the SFHo-18.6 SN simulation.  Note that reabsorption becomes less relevant at smaller values of $|g_{a\gamma\gamma}|$.
    }
    \label{fig:axion_prod_spec_trapping}
\end{figure}

Importantly, we account for the reabsorption of axions inside the PNS through inverse nucleon bremsstrahlung and pion conversion, following the method described in Ref.~\cite{Caputo:2021rux,Lella:2023bfb}. Note that in a PNS only a large number of $\pi^-$ is expected, as opposed to $\pi^+$, and therefore the main production mechanism through pion conversion is $\pi^-\, p \to n\, a$. For axion reabsorption we account for $a\, p \to \pi^+ \, n$, $a\, p(n) \to \pi^0\, p(n)$, and $a\, p(n) \to \pi^+\, n(p)$. The number of axions per unit  energy and unit time is given by
\begin{equation}
\frac{d^2 N_a}{d E_a d t}=\int_0^{\infty} 4 \pi r^2 d r\left\langle e^{-\tau\left(E_a^*, r\right)}\right\rangle \frac{d^2 n_a}{d E_a d t} \,,
\end{equation}
with the angle-averaged absorption coefficient is \cite{Caputo:2022rca}
\begin{align}
\left\langle e^{-\tau\left(E_a^*, r\right)}\right\rangle=\frac{1}{2} \int_{-1}^{+1} d \mu e^{-\int_0^{\infty} d s \Gamma_a\left(E_a^*, \sqrt{r^2+s^2+2 r s \mu}\right)} \,.
\label{eq:absorption_coeff}
\end{align}
Here we define the reduced absorption rate 
\begin{align}
\Gamma_a\left(E_a, r\right)=\lambda_a^{-1}\left(E_a, r\right)\left[1-e^{-\frac{E_a}{T(r)}}\right]  \,,
\end{align}
where $\lambda_a$ is the axion mean free path and $T(r)$ is the PNS temperature. Note that \eqref{eq:absorption_coeff} is evaluated at $E_a^*=E_a \alpha(r) / \alpha\left(\sqrt{r^2+s^2+2 r s \mu}\right)$ with $\alpha(r)$ the gravitational lapse parameter. In Fig.~\ref{fig:axion_prod_spec_trapping} we show the axion production spectrum with and without accounting for reabsorption within the PNS for $g_{a\gamma\gamma}=10^{-11}\, \mathrm{GeV}^{-1}$, assuming the GUT benchmark model.

\section{Supernovae analysis Additional Details}~\label{SM:additional_analyis_details}

In this Section we discuss in more detail our analyses of SMM and Fermi-LAT data towards various SN. To begin, let us comment on the regions of the $(m_a,|g_{a\gamma\gamma}|)$ plane excluded by these analyses. Note that at sufficiently low axion masses the axion decay flux is proportional to $g_{a\gamma\gamma}^4m_a^{2}$
, such that the upper limit on the axion-photon coupling scales as $|g_{a\gamma\gamma}| \propto m_a^{-\frac{1}{2}}$. 
If the observation time is sufficiently long after the SN, such that the signal may exponentially decay before reaching the detector at large $g_{a\gamma\gamma}$, Eq.~\eqref{eq:differential_flux} implies that the upper boundary of the excluded region scales as $|g_{a\gamma\gamma}| \propto m_a^{-1}$ . This is the case for {\it e.g.} the Fermi observations of SN2024ggi, SN2023ixf, and SN1987A. 

For smaller delay times, as in the case of the SMM observation of SN1987A, the signal flux is still suppressed if axions mostly decay inside the SN photosphere, {\it i.e.} if the axion decay length $d_a = \tau_a \gamma_a \beta_a < R_\mathrm{ps}$, with $R_\mathrm{ps}$ the photosphere radius, $\beta_a$ the axion velocity, and $\gamma_a$ the Lorentz factor. In this scenario the upper boundary scales as  $|g_{a\gamma\gamma}| \propto m_a^{-2}$.
In our Fermi-LAT analyses of SN2023ixf and SN2024ggi, and our SMM analysis of SN1987A, at even lower $m_a$ along this boundary the sensitivity drops off exponentially due to the reabsorption of axions inside the PNS, such that the constraint becomes approximately flat as a function of $|g_{a\gamma\gamma}|$. Note that the sensitivity for the SMM constraint extends to higher $|g_{a\gamma\gamma}|$ compared to the Fermi-LAT bounds as the SMM analysis is only sensitive to photon energies $E_\gamma < 100 \, \mathrm{MeV}$ and thus lower energy axions, for which reabsorption is less efficient.   

In cases where the beginning of the observation window occurs after the SN onset, there is a nonzero minimal delay time  $t_\mathrm{min}$ between the optical signal from line-of-sight photons  and the axion decay signal. This is the case for the Fermi-LAT observations of SN1987A and Cas A, with minimal delay times of $21$ and $\sim 330$ years, respectively, given that our Fermi data sets begin shortly after the Fermi launch date in 2008. Consequently for these SN, the requirement that the photon resulting from the axion decay arrives with the energy $E_\gamma$ and a delay time which is at least $t_\mathrm{min}$ enforces that \cite{Muller:2023vjm}
\begin{align}
t_\mathrm{min}<\frac{2 m_a^2}{4 E_\gamma^2-m_a^2} R_{\mathrm{SN}} 
\end{align}
for $E_\gamma>\frac{m_a}{2}$ (for  $E_\gamma \le \frac{m_a}{2}$ there is no constraint on $t_\mathrm{min}$). This is because the photon's energy in the lab frame $E_\gamma$ is related to its energy in the axion's rest frame, $m_a/2$, by  $m_a/2E_\gamma = \gamma_a\left(1-\beta_a\cos\alpha\right)$, where $\alpha$ is the angle between the momenta of the axion and the photon. Then, if $E_\gamma>\frac{m_a}{2}$, the photon must have been emitted with a positive projection onto the axion's propagation direction, and in particular $\cos\alpha >\sqrt{1-\frac{m^2_a}{4E^2_\gamma}}$. As a result, there is a maximal time delay for the photon to reach the detector, which due to the slowness of the axion in this region of the parameter space corresponds to the limit where the axion decays right at the detector. However, if $E_\gamma<\frac{m_a}{2}$, the axion could have traveled away form the detector and decayed infinitely far, thus resulting with an arbitrarily large time delay of the photon signal. 

As our Fermi-LAT data sets do not extend below $100 \, \mathrm{MeV}$, this relation implies that our analyses have no sensitivity for $m_a < 1.6 \, \mathrm{MeV}$ ($m_a < 23 \, \mathrm{MeV}$ ) for SN1987A (Cas A). Recall that even though SN2023ixf and SN2024ggi went off while Fermi was observing, for these SN $t_\mathrm{min}$ is  effectively still nonzero due to our lack of knowledge of the precise time when the SN occurred. As discussed in the main text, to be conservative we assume that Fermi's effective area is zero during one orbital rotation of the satellite such that $t_\mathrm{min}=90$ minutes. This implies that the SN2023ixf and SN2024ggi analyses constrain $m_a \ge 0.4 \, \mathrm{keV}$.

\subsection{SMM Analysis of SN1987A}
\label{sec:smm_analysis}

The flux of decaying axions from SN1987A in the energy range $[25,100] \,\mathrm{MeV}$, during the first $223$ seconds after onset, is constrained by SMM data to be $F_\gamma < 1.19 /\mathrm{cm}^2 $ ($1.78 /\mathrm{cm}^2 $) at $1\sigma$ ($2\sigma$) 
\cite{Jaeckel:2017tud}. Note that since  SMM observed SN1987A for only 223 seconds, the computation of the signal flux is greatly simplified compared to \eqref{eq:differential_flux} as we may assume a small angle approximation to reduce the integral of the flux over time, photon energy, and angle to a one-dimensional integral over axion energy \cite{Muller:2023pip}:
\begin{align}
& F_\gamma =\frac{\mathrm{Br}_{a\to\gamma\gamma}}{2 \pi R_{\mathrm{SN}}^2} \int_{m_a}^{\infty} \mathrm{d} E_a\left[\Delta E_\gamma\left(p_a\right) e^{-\frac{R_\mathrm{ps} m_a}{\tau_a p_a}}-\frac{\tau_a}{t^{\max }} \frac{m_a}{2}\left(e^{-\frac{t^{\max }}{\tau_a} \frac{2 E_\gamma^{\min }\left(p_a\right)}{m_a}}-e^{-\frac{t^{\max }}{\tau_a} \frac{2 E_\gamma^{\max }\left(p_a\right)}{m_a}}\right)\right] p_a^{-1} \frac{\mathrm{~d} N_a}{\mathrm{~d} E_a} \Theta\left(\Delta E_\gamma\left(p_a\right)\right),
\end{align}
where $\mathrm{Br}_{a\to\gamma\gamma}$ is the branching ratio of axion decays into photons, $t_\mathrm{max}=223$ s, and with
\begin{align}
E_\gamma^{\min }\left(p_a\right) & =\max \left(25 \,\mathrm{MeV}, \frac{1}{2}\left(E_a-p_a\right), \frac{m_a^2 R_\mathrm{ps}}{2 p_a t^{\max }}\right), \\
E_\gamma^{\max }\left(p_a\right) & =\min \left(100 \,\mathrm{MeV}, \frac{1}{2}\left(E_a+p_a\right)\right), \\
\Delta E_\gamma\left(p_a\right) & =E_\gamma^{\max }\left(p_a\right)-E_\gamma^{\min }\left(p_a\right) \,.
\end{align}
Our constraint on the axion-photon coupling from the SMM data for the GUT benchmark model is shown in Fig. \ref{fig:gayy_constraints}.

\subsection{Fermi-LAT Analysis Extended Details for SN1987A}
\label{sec:fermi_analysis_SN1987A}

Axions produced in SN1987A may be detected by Fermi, which began taking data in 2008, if the lifetime of the axion is larger than $\mathcal{O}(10)$ years. 
Note that the signal is spread over angles $\theta \sim \mathcal{O}(t/R_\mathrm{SN}) \sim 10^{-4}$ and is therefore detected by Fermi-LAT as a PS. However, the small-angle approximation we use for the SMM analysis of SN1987A is not valid in this case and the flux should be integrated directly over time and angle.\footnote{As the integrand typically has a highly localized support in the $(\theta, t, E_\gamma)$ plane, we use the \texttt{vegas} integration package (in \textit{batch mode}), which implements the Monte Carlo algorithm of Ref.~ \cite{Lepage:2020tgj}.}

To assess the importance of the finite energy resolution of Fermi-LAT in our analysis, we convolve the signal flux
\begin{equation}
\mu^\mathrm{sig}_i \to\sum_j \mu^\mathrm{sig}_j \cdot G\left(E_i, E_j\right) \cdot \Delta E_j
\end{equation}
(with $\Delta E_j = E_{j+1}-E_j$) with the Gaussian kernel
\begin{equation}
G\left(E^{\prime}, E\right)=\frac{1}{\sqrt{2 \pi} \sigma(E)} \exp \left(-\frac{\left(E^{\prime}-E\right)^2}{2 \sigma(E)^2}\right) \,,
\end{equation}
where we fix $\sigma(E) = 0.15 E$, which is approximately the energy dispersion of the Fermi-LAT data in the 100 MeV - 1 GeV range. 
We verify that this procedure gives only sub-percent level changes to our  constraints from SN1987A. The effect is similarly negligible for the other Fermi-LAT analyses performed in this work.

\subsection{Fermi-LAT Analysis Extended Details for Cas A}
\label{subsectio:Cas_A}

Compared to the other SN studied in this work, the maximal angular spread of the axion decay signal from Cas A for excluded $g_{a\gamma\gamma}$ is the largest at $\mathcal{O}(10)$ degrees, which can be resolved by Fermi-LAT. Indeed, Cas A is the closest SN to Earth at $3.4$ kpc and the minimal time delay probed by Fermi-LAT is the largest at $\sim 330$ years. In Fig. \ref{fig:Cas_A_angular_spread} we show the 68\% containment radius of the signal as a function of gamma-ray energy for various axion masses.

To estimate the 95\% upper limit on $g_{a\gamma\gamma}$ from our Fermi-LAT data, we may first assume that the 90\% containment radius of the signal is 3 degrees, resulting in an upper limit $|g_{a\gamma\gamma}|^{95,*}(m_a)$.  
We may then refine this bound by performing a spatial template analysis, for each $m_a$, assuming a 90\% containment radius equal to that of the axion decay signal for $|g_{a\gamma\gamma}|=|g_{a\gamma\gamma}|^{95,*}(m_a)$. We perform the analogous procedure for the upper boundary of the exclusion region, which results in the constraint shown in Fig. \ref{fig:gayy_constraints}.

\section{Other heavy axion supernovae constraints}
\label{sec:other}

In this Section we briefly discuss existing SN constraints on heavy axions in the $10$ keV - $1$ GeV range. We begin by considering existing constraints on the axion-photon coupling, notably from SN1987A, and then discuss how these same constraints are modified for the GUT benchmark model which has axion-nucleon and axion-pion couplings. 

\subsection{Axions with no gluon coupling}

Most previous works using SN to constrain heavy axions \cite{Jaeckel:2017tud,Hoof:2022xbe,Muller:2023vjm} assume axion production in the SN occurs only through the axion-photon coupling, which allows emission through the Primakoff process and, for $m_a \ge 2 \omega_\mathrm{pl}$, photon coalescence $\gamma\gamma \to a$. The constraints which result are relevant for axions that do not couple to gluons, though as we show below, even in the case where the axion only couples to electroweak gauge bosons in the UV, additional production channels can be relevant as the axion acquires loop-induced couplings to nucleons (see Sec. \ref{sec:RGflow}).

Existing SN constraints on the axion-photon parameter space for heavy axions include the following (see Fig.~\ref{fig:limits_summary_plot_photon_couplings} for a summary). Ref. \cite{Muller:2023vjm} set a constraint on decaying heavy axions (not coupling to gluons) from the SMM observation of SN1987A, using the same formalism for the photon signal flux as in this work. Note that the excluded parameter space is actually not contiguous, as pointed out in Ref. \cite{Diamond:2023scc}. For large enough $g_{a\gamma\gamma}$, the photons resulting from decaying axions produced by the SN form a gas, which can reach densities sufficiently large to trap the photons by pair production, creating a QED plasma. In this case the typical energies of the photons are reduced from $\mathcal{O}(100)$ MeV to $\mathcal{O}(100)$ keV, outside of the range of the SMM detector. On the other hand, this portion of parameter space for which the SMM bound does not apply is  excluded by the non-observation of the reprocessed lower energy photons in the Pioneer Venus Orbiter, as shown in \cite{Diamond:2023scc}.

Refs.~\cite{Caputo:2021rux,Lucente:2020whw} looked at the axion emission rate in a SN explosion, which is constrained by the neutrino burst observed for SN1987A. Ref.~\cite{Caputo:2022mah} obtained bounds on $|g_{a\gamma\gamma}|$ by studying core-collapse SN with low explosion energy, which constrain the energy deposition between the collapsing core and the surface of the progenitor star due to axion decays.  On the other hand, Ref.~\cite{Caputo:2021rux} studied the diffuse axion background from all past SN, establishing constraints from its contribution to the cosmic photon background. Lastly, Ref. \cite{Diamond:2023cto} established a bound on axions in the 1 to 400 MeV range from X-ray observations of the NS merger GW170817, as axions are produced in the merger and may radiatively decay into $\mathcal{O}(100)$ MeV photons.
Note that in addition to those from SN,  constraints on heavy axions arise from the decay of the irreducible freeze-in axion relic density \cite{Langhoff:2022bij}, from BBN \cite{Baumholzer:2020hvx} (in Fig. \ref{fig:limits_summary_plot_photon_couplings} we show their bound which assumes a reheating temperature of $10 \, \mathrm{MeV}$), and from NuSTAR observations of the M82 starburst galaxy \cite{Candon:2024eah}.

We may compare these existing constraints to those from our Fermi-LAT data sets towards SN2023ixf and SN2024ggi, assuming the axion does not couple to gluons, but accounting for production both through the axion-photon interaction and through nucleon couplings that are generated in the IR through RG-flow.  The result is shown in Fig. \ref{fig:limits_summary_plot_photon_couplings}, where for concreteness we assume the axion couples with equal strength to the electroweak bosons, {\it i.e.}, $C_{aWW}=C_{aBB}$ (see~\cite{Manzari:2024jns}). In this scenario, the Fermi-LAT data toward SN1987A and Cas A do not give any constraints, as the relevant $|g_{a\gamma\gamma}|$ values result in the decay of axions between SN and the beginning of Fermi-LAT. 
Note that our bound on heavy axion decays from SN2023ixf and SN2024ggi is significantly stronger than the previous one using Fermi-LAT data towards SN2023ixf of Ref.~\cite{Muller:2023pip} (labeled `SN2023ixf' in Fig. \ref{fig:limits_summary_plot_photon_couplings}). The improvement is largely due to a more appropriate set of data quality cuts (see Section \ref{sec:data_reduction}), which removed regions of vanishing effective area in the analysis of ~\cite{Muller:2023pip}. The bound is further strengthened, though to a much lesser extent, by accounting for the additional axion production through pion conversion and nucleon bremsstrahlung at the 1-loop level.

\subsection{Axions with gluon couplings}

Let us now discuss how the existing constraints on $|g_{a\gamma\gamma}|$ are affected by turning on the axion coupling with the gluons. First, the freeze-in bound, assuming a minimal reheat temperature $T_\mathrm{RH} = 5 \,\mathrm{MeV}$, is unaffected because of the small baryon-to-photon ratio. Moreover, we recast the other SN bounds previously discussed, including all relevant production and decay processes in the GUT scenario. This includes the ``cooling'' bound obtained in Ref.~\cite{Caputo:2021rux}, the low energy SN bound obtained in Ref.~\cite{Caputo:2022mah}, and the diffuse axion background obtained in Ref.~\cite{Caputo:2021rux}. We follow the procedures outlined in the references above with the only difference being that the axion couples to gauge bosons in a GUT symmetric manner, as described in the main text of this work.  
Two main results emerge from our analysis. In the free-streaming regime, a stronger bound on $g_{a\gamma\gamma}$ arises due to the enhanced axion emissivity caused by the high density of nucleons in the PNS. Furthermore, the increased axion scattering rate with nucleons and pions shifts the onset of trapping effects to lower values of $g_{a\gamma\gamma}$.

\begin{figure}
    \centering
    \includegraphics[width=0.6\linewidth]{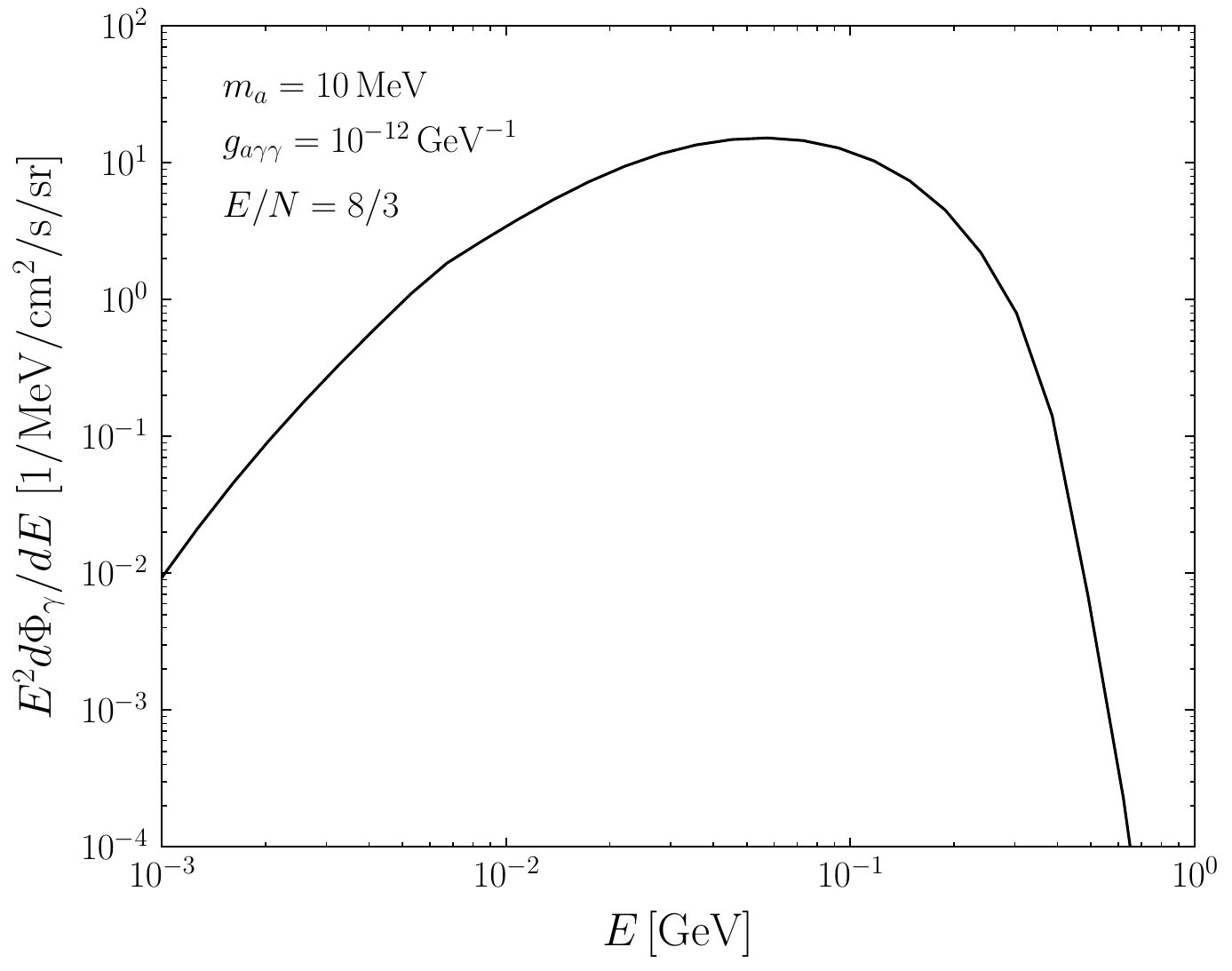}
    \caption{The diffuse gamma-ray flux as a function of photon energy $\omega$, from decays of the cosmic axion background for $m_a = 10 \,\mathrm{MeV}$, $g_{a\gamma\gamma}=10^{-12} \, \mathrm{GeV}^{-1}$, and with $E/N = 8/3$, as in our benchmark GUT model.}
    \label{fig:diffuse_ALP}
\end{figure}

The bound from a diffuse axion background turns out to be relevant in the parameter space of interest for this work (`Diffuse $\gamma$-ray' in Fig.~\ref{fig:gayy_constraints}), and therefore we outline here the details of its derivation. Axion emission from all past SN generates a cosmic axion background density. The radiative decays of these axions contribute to the diffuse cosmic gamma-ray background, which can thus be constrained. Closely following Refs.~\cite{Caputo:2021rux, Vitagliano:2019yzm}, the density of photons (from axion decays) today is
\begin{align}
    \frac{d n_{\gamma}}{dE} = \int_{0}^{\infty} dz (1+z)n^{\prime}_\mathrm{cc}(z)\int_{E_{z, \mathrm{min}}}^{\infty}d E_z f_D(E_z)\frac{2}{p_a(E_z)}F_a(E_z)\,,
\end{align}
where $p_a(E_z)=\sqrt{E_z^2-m_a^2}$, $E_{z, \mathrm{min}} = (E' + m_a^2/4E')$ with $E'=E(1+z)$, and $F_a(E) = dN_a/dE$ is the axion spectrum emitted by an average SN, for which we use model SFHo-18.6. We also define the collapse number per comoving volume per redshift interval $n^{\prime}_\mathrm{cc}(z)$, and the fraction of axions that have decayed, outside the stellar photosphere, between the epoch of emission at redshift $z$ and today
\begin{align}
    f_D(E_z) = {\rm exp}\bigg(-\frac{R_\mathrm{ps}}{\tau_{a}}\frac{m_a}{p_a(E_z)}\bigg) - {\rm exp}\bigg[-\int_{0}^z d z_D \frac{t^{\prime}(z_D)}{\tau_a} \frac{m_a}{p_a(E_z)}\frac{1+z}{1+z_D}\bigg] \,.
\end{align}
Finally, we use $n^{\prime}_\mathrm{cc} (z) = R_\mathrm{cc}(z)t^{\prime}(z)$, where $t^{\prime}(z) = dt/dz = \bigg(H_0(1+z)\sqrt{\Omega_M(1+z)^3+\Omega_{\Lambda}}\bigg)^{-1}$ and $R_\mathrm{cc}(z)$ is the cosmic core-collapse rate. For the latter, we use the most conservative (lowest) rate discussed in Ref.~\cite{Caputo:2021rux}, {\it i.e.} the analytic approximation of Ref.~\cite{Madau:2014bja}. The photon spectrum obtained for $m_a = 10 \,\mathrm{MeV}$ and $g_{a\gamma\gamma}=10^{-12} \, \mathrm{GeV}^{-1}$ is shown in Fig.~\ref{fig:diffuse_ALP}. We compare this result with the extragalactic gamma-ray flux reported in Ref.~\cite{Fermi-LAT:2014ryh}. In particular we note that the measured flux in the energy range $1$ MeV $\--$ $1$ GeV
is approximately flat and
\begin{align}
    E^2\frac{d\Phi_{\gamma}}{dE} \approx 2\times 10^{-3}\; {\rm MeV\; cm^{-2}\; s^{-1}\; sr^{-1}}\,,
\end{align}
with $d \Phi_\gamma / d E=\left(d n_\gamma / d E\right) / 4 \pi$. We require the axion-induced flux to be less than the measured flux above at all energies. 
Finally, we note that while we are able to reproduce the results in Ref.~\cite{Caputo:2021rux} for the case of an axion only coupled to photons, we are unable to reproduce the results in Ref.~\cite{Lella:2024dmx}, which consider also tree-level couplings to quarks and gluons. Our limits are stronger by about one order of magnitude relative to those in Ref.~\cite{Lella:2024dmx}.

\end{document}